\documentclass[]{pasj02} 
\usepackage[switch,mathlines]{lineno} 
\usepackage{url} 

\jyear{2025}
\Received{}
\Accepted{}

\usepackage{bm,color,xcolor}
\usepackage{fix-cm}
\usepackage{ulem} 

\begin{document} 

\newcommand{\gw}{S240422ed}
\newcommand{\Ks}{$K_{\rm s}$}
\newcommand{\numofmoircsobsgalaxies}{206}
\newcommand{\numoftransientsobswithmoircs}{8}
\newcommand{\thresigmaforcounterpart}{5}
\newcommand{\sigmaforvarithre}{5}
\newcommand{\completeness}{22}

\title{ 
J-GEM near-infrared follow-up observations of the gravitational wave event S240422ed
}

\author{
Ichiro \textsc{Takahashi}\altaffilmark{1}\altemailmark\orcid{0000-0003-2691-4444},
Tomoki \textsc{Morokuma}\altaffilmark{2}\orcid{0000-0001-7449-4814}, 
Masaomi \textsc{Tanaka}\altaffilmark{3,4}\orcid{0000-0001-8253-6850},
Mahito \textsc{Sasada}\altaffilmark{5,6}\orcid{0000-0001-5946-9960}, 
Hiroshi \textsc{Akitaya}\altaffilmark{2,6,7}\orcid{0000-0001-6156-238X},
Ichi \textsc{Tanaka}\altaffilmark{8}\orcid{0000-0002-4937-4738},
Nozomu \textsc{Tominaga}\altaffilmark{9,10,11}\orcid{0000-0001-8537-3153},
Michitoshi \textsc{Yoshida}\altaffilmark{6,9,10}\orcid{0000-0002-9948-1646},
Yousuke \textsc{Utsumi}\altaffilmark{9,12}\orcid{0000-0001-6161-8988},
Ryosuke \textsc{Itoh}\altaffilmark{6,13}\orcid{0000-0002-1183-8955},
Kyohei \textsc{Kawaguchi}\altaffilmark{14,15}\orcid{0000-0003-4443-6984},
and the J-GEM collaboration
}
\altaffiltext{1}{Department of Physics, Institute of Science Tokyo, 2-12-1 Ookayama, Meguro-ku, Tokyo 152-8550, Japan}
\altaffiltext{2}{Astronomy Research Center, Chiba Institute of Technology, 2-17-1, Tsudanuma, Narashino, Chiba 275-0016, Japan}
\altaffiltext{3}{Astronomical Institute, Tohoku University, Sendai, Miyagi 980-8578, Japan}
\altaffiltext{4}{Division for the Establishment of Frontier Sciences, Organization for Advanced Studies, Tohoku University, Sendai 980-8577, Japan}
\altaffiltext{5}{Institute of Integrated Research, Institute of Science Tokyo, 2-12-1 Ookayama, Meguro-ku, Tokyo 152-8550, Japan}
\altaffiltext{6}{Hiroshima Astrophysical Science Center, Hiroshima University, Kagamiyama 1-3-1, Higashi-Hiroshima, Hiroshima 739-8526, Japan}
\altaffiltext{7}{Planetary Exploration Research Center, Chiba Institute of Technology, 2-17-1, Tsudanuma, Narashino, Chiba 275-0016, Japan}
\altaffiltext{8}{Subaru Telescope, National Astronomical Observatory of Japan, National Institutes of Natural Sciences, 650 North A'ohoku Place, Hilo, HI 96720, USA}
\altaffiltext{9}{National Astronomical Observatory of Japan, 2-21-1 Osawa, Mitaka, Tokyo 181-8588, Japan}
\altaffiltext{10}{Astronomical Science Program, The Graduate University for Advanced Studies (SOKENDAI), 2-21-1 Osawa, Mitaka, Tokyo 181-8588, Japan}
\altaffiltext{11}{Department of Physics, Konan University, 8-9-1 Okamoto, Kobe, Hyogo 658-8501, Japan}
\altaffiltext{12}{Vera C.\ Rubin Observatory, Avenida Juan Cisternas \#1500, La Serena, Chile}
\altaffiltext{13}{Bisei Astronomical Observatory, Ibara, Okayama 714-1411, Japan}
\altaffiltext{14}{Max Planck Institute for Gravitational Physics (Albert Einstein Institute), Am M\"{u}hlenberg 1, Potsdam-Golm, 14476, Germany}
\altaffiltext{15}{Center of Gravitational Physics and Quantum Information,
 Yukawa Institute for Theoretical Physics, Kyoto University, Kyoto, 606-8502, Japan}

\email{itakahashi@astro.phys.sci.isct.ac.jp}
%


\maketitle

\begin{abstract}
We report our near-infrared (NIR) follow-up observations of the gravitational wave (GW) event S240422ed using the Subaru Telescope/MOIRCS. S240422ed was initially classified as a black hole-neutron star merger with $>$ 99\% probability of electromagnetic wave emission. We started follow-up observations 7.8 hours after the event. Over two nights, we observed \numofmoircsobsgalaxies\ nearby galaxies in $Y$ and {\Ks} bands down to about 21.4 and 21.1 AB mag (3$\sigma$), respectively. The total completeness of our survey based on galaxy $B$-band luminosity is \completeness\%. As a result of our observations, five candidate counterparts were identified. We show that properties of these five objects are not consistent with kilonova such as AT2017gfo. Four objects are consistent with known classes of transients such as supernovae or dwarf nova outbursts. On the other hand, the nature of the remaining one object, which shows a red color and rapid decline, remains unclear. Although later analyses of GW signal reclassified S240422ed as likely terrestrial noise, our NIR observations provide valuable lessons for future NIR surveys for GW sources. We demonstrate that deep NIR follow-up observations as presented in this work would effectively constrain the presence of red kilonova even at 200 Mpc distance. We also discuss the importance of deep and wide NIR reference images and of understanding the properties and frequency of Galactic transients.
\end{abstract}


\section{Introduction}

Compact binary mergers are important phenomena in a wide area of astrophysics.
At the coalescence, they emit strong gravitational waves (GWs), which are one of the main targets of ground-based GW detectors, such as LIGO, Virgo and KAGRA \citep{ligo15,virgo15,kagra21}.
When the binary mergers include at least one neutron star (NS), i.e., binary NS mergers or black hole (BH)-NS mergers, they may produce short gamma-ray bursts by launching the relativistic jets.
In addition, such events are also one of the promising sites for rapid neutron capture nucleosynthesis ($r$-process, e.g., \cite{lattimer74,eichler89,goriely11,korobkin12,bauswein13,wanajo14}).

The first GW from a binary NS merger was observed in 2017 (GW170817, \cite{abbott17}).
For this event, various electromagnetic (EM) emissions were also observed \citep{abbott17MMA}, including a short gamma-ray burst, subsequent afterglow in X-ray and radio wavelengths, and thermal emission in the ultraviolet-optical-infrared wavelengths (known as AT2017gfo).
In particular, intensive follow-up observations of AT2017gfo were conducted (e.g., \cite{arcavi17,chornock17,coulter17,cowperthwaite17,muccully17,shappee17,smartt17,tanvir17,Utsumi2017}), revealing that the properties of this transient are broadly consistent with a kilonova, thermal emission powered by radioactive decay of $r$-process nuclei (\cite{li98,metzger10}).
Combining these observations with theoretical modeling,
it was confirmed that NS merger is a site of $r$-process nucleosynthesis (e.g., \cite{kasen17,perego17,tanaka17,rosswog18,kawaguchi18}).

Since then, several efforts have been made to observe kilonovae associated with GW events.
For example, the second GW detection of binary NS merger event GW190425 \citep{abbott20} and promising BH-NS merger events (GW200105 and GW200115, \cite{abbott21}) were reported. 
However, due to the poor localization of these GW events, no EM counterpart was discovered despite the large efforts (see \cite{coughlin19,hosseinzadeh19,paek24,coulter2025ApJ} for GW190425, and \cite{anand21,dichiara21} for GW200105/GW200115).

To date, there is only one kilonova observed in association with GW detection, and thus, the general picture of $r$-process nucleosynthesis in binary NS mergers and BH-NS mergers is not yet clear.
In fact, depending on the mass and mass ratio of the merging objects, a large diversity in the mass ejection and nucleosynthesis is expected (e.g., \cite{radice20,fujibayashi23}).
Accordingly, properties of kilonovae can also have a large variety (e.g., \cite{kawaguchi20,just23}).
To understand such variety in the mass ejection and nucleosynthesis, it is important to observe more kilonovae from a variety of binary NS mergers and BH-NS mergers.
GW signals provide information of mass and mass ratio of merging objects, while EM signals provide information of ejected material. 
Thus, multi-messenger observations of both GW and EM signals are essential.

LIGO-Virgo-KAGRA (LVK) collaboration started the second part of the 4th observing (O4b) run on 2024 April 10.
In this paper, we present our follow-up observations for \gw\ (see also \cite{EMfollowup_WINTER_2025, EMfollowups_Pillas2025,EMfollowup_GW-MMADS_2025} for follow-up observations of GW events during O4).
\gw\ was detected at UT 21:35:13.417 on April 22, 2024 \citep{S240422ed_gcn1}.
From the real-time GW data analysis with {\tt GstLAL} pipeline, the false alarm rate was estimated to be $3.1 \times 10^{-13}$ Hz or about one in $10^5$ years.
The category of the event was estimated to be a BH-NS merger with a probability of $> 99\%$.
The probability of having a remnant around the central object (HasRemnant parameter), which indicates the likelihood of EM emission, was also $> 99\%$.
Furthermore, this event was also localized to a $90\%$ probability region of 441 deg$^2$ (with a distance of $214 \pm 64$ Mpc).
These measurements motivated EM follow-up observations with optical/near-infrared (NIR) telescopes all over the world.
However, with an updated GW data analysis, the classification of \gw\ was largely altered about two months later (2024 July 3, \cite{S240422ed_gcn3}):
93\% probability for terrestrial noise,
5\% probability for binary NS merger,
and 2\% probability for BH-NS merger
(false alarm rate of $3.269 \times 10^{-7}$ Hz or about one in 35 days).

Our collaboration, Japanese collaboration for Gravitational wave ElectroMagnetic follow-up (J-GEM; \cite{morokuma2016,Yoshida2017,tominaga18,Sasada:2021tn}), performed an EM counterpart search for \gw\ targeting nearby galaxies with an NIR camera (MOIRCS: Multi-Object InfraRed Camera and Spectrograph, \cite{suzuki08,ichikawa06}) mounted on the Subaru Telescope 7.8~hours after the estimated time of the merger \citep{morokuma24a}.
Despite the high probability that \gw\ is of non-astrophysical origin, our observations provide a valuable test case for deep NIR follow-up observations targeting the nearby galaxies.

We describe our observations and data reduction in section~\ref{sec:observations}, and present results of our transient search in section~\ref{sec:results}.
Then, we discuss the nature of detected candidates and lessons from our observations in section~\ref{sec:discussion}.
Finally, we give a summary of our paper in section~\ref{sec:conclusions}.
Throughout this paper, all the magnitudes are given in AB magnitudes and all the times are given in UT.

\section{Observations and data analysis}
\label{sec:observations}
\subsection{Subaru/MOIRCS observations}
\label{sec:obs}
We conducted NIR imaging observations of $\numofmoircsobsgalaxies$ nearby galaxies with MOIRCS on the Subaru Telescope in the first-half nights of 2024 April 23 and 24 \citep{morokuma24a,Morokuma:2024ab}. 
Following the galaxy-targeted survey strategy described in \citet{Sasada:2021tn}, we selected galaxies from the GLADE catalog version 2.4\footnote{In the GLADE catalog, distances are calculated using a flat $\Lambda$CDM cosmology with the following parameters: $H_0 = 100h = 70\ \mathrm{km\ s^{-1}\ Mpc^{-1}}$, $\Omega_M = 0.27$, and $\Omega_\Lambda = 0.73$.} \citep{GLADE2018} based on the three-dimensional (3D) localization map of the GW event.  
We prioritized galaxies according to their weighted probabilities, calculated from the 3D localization probability and $B$-band luminosity.
Galaxies were selected in descending order of the weighted probability, subject to observability constraints from the Subaru Telescope during the observing nights.  
No distance cutoff was applied.

The high GW probability region is located in the southern hemisphere and the target visibility is not optimal for the Subaru telescope.
Due to the limitation of observing time, we decided to take one exposure in $Y$ and {\Ks} bands, respectively, in a night for a given galaxy, resulting in two exposures in the two filters in two nights. 
Multiple exposures enable us to find moving objects such as asteroids which are one of the major contaminants in search for extragalactic transients. 
The exposure times are 15 and 20~sec in $Y$-band and 16 and 20~sec in {\Ks}-band. 
On the second night, we also took NIR images for \numoftransientsobswithmoircs\ possible counterparts reported by other groups \citep{Hu2024GCN.36273....1H, Kumar:2024aa}. 
The observations typically reached 3$\sigma$ limiting magnitudes of about 21.4~mag in $Y$ and 21.1~mag in {\Ks} bands, estimated from the signal-to-noise ratio (S/N) of isolated point sources. The typical full widths at half maximum (FWHMs) of the point spread function (PSF) were 0.9~arcsec in $Y$ and 0.7~arcsec in {\Ks} bands.
All observations were conducted under stable weather conditions, and both the telescope and the instrument operated without any issues throughout the observing nights.
Sky locations of the observed galaxies are shown in figure~\ref{fig:skymap}. 
A complete list of the observed galaxies is provided in appendix~\ref{sec:appendix_gallist}.

\begin{figure}[ht]
  \begin{center}
     \includegraphics[width=\columnwidth]{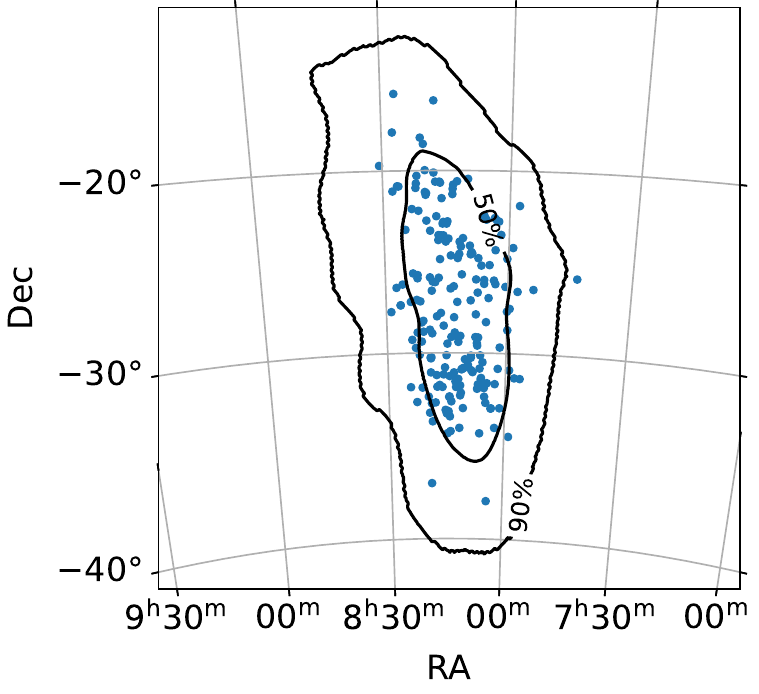}
  \end{center}
  \caption{
  Sky localization map of the gravitational wave event S240422ed, overlaid with the positions of galaxies targeted in our follow-up observations. The background shows the probability density map derived from the 3D localization of the gravitational wave. The 50\% and 90\% confidence regions are indicated by labeled solid contours. The positions of the galaxies observed in our follow-up are shown as blue scatter points.
  {Alt text: Sky map of a gravitational wave event showing probability contours and observed galaxy positions. The horizontal and vertical axes represent right ascension and declination in equatorial coordinates.}
  }
  \label{fig:skymap}
\end{figure}

The MOIRCS data were reduced in a standard way using the MCSRED2 data reduction package\footnote{MCSRED2 data reduction package $\langle$\url{https://www.naoj.org/staff/ichi/MCSRED/mcsred_e.html}$\rangle$.}. 
The WCS solution was calculated using the \texttt{astrometry.net} software. 
The distances of the observed galaxies range from 116 to 339~Mpc, with a median of 184~Mpc and a standard deviation of 33~Mpc, corresponding to an angular diameter distance of approximately 1~kpc arcsec$^{-1}$.  
The apparent sizes of these galaxies are much smaller than the field-of-view of each of the two MOIRCS detectors ($4 \times 7$~arcmin$^{2}$).
Thus, we subtracted the background by performing a median sky subtraction using the three frames before and after, and a two-dimensional low-order fit to the image.

Flux calibration was done using multiple catalogs, depending on the sky location and filter. 
The public catalogs from the VISTA surveys (VISTA Hemisphere Survey, VHS, \cite{McMahon2013}; and VISTA Variables in the Vía Láctea eXtended Survey, VVVX, \cite{saito24A&A}) and from the Two Micron All Sky Survey (2MASS; \cite{Skrutskie2006AJ....131.1163S}) were used for the {\Ks}-band data 
while the Pan-STARRS1 (PS1; \cite{flewelling20}) and the Dark Energy Camera Plane Survey 2 (DECaPS2; \cite{waters20}) catalogs were used for the $Y$-band data.

\subsection{Survey completeness}
\label{sec:completeness}
We evaluated the survey completeness of our follow-up observations.
It is defined as the sum of the probabilities that each observed galaxy is the host of the GW event.
This definition effectively quantifies the coverage of host-galaxy candidates in our follow-up observations.
This contrasts with volume-based completeness estimates in wide-field surveys.

The GW alert gives a 3D localization map: a probability $p$ is assigned at each 3D position of the sky $\bm{r}$. 
We used the 3D localization map provided by BAYESTAR \citep{S240422ed_gcn1} and Bilby \citep{S240422ed_gcn2}.
The compact binary merger rate may also depend on the stellar mass or star formation rate of the galaxy. 
Thus, we weighted the probability with the properties of the galaxy. A possible choice of the weight includes (i) $K$-band luminosity of the galaxy, which is a proxy of the stellar mass, (ii) $B$-band luminosity of the galaxy, which is a proxy of star formation rate, or (iii) a combination of these two. In our work, since the $K$-band luminosity is available only for $\sim$60\% of the galaxies in the GLADE catalog, we weighted the probability with $B$-band luminosity ($L_B$, see also \cite{Sasada:2021tn}).

Namely, we define the weighted probability $P^i$ for galaxy $i$ as follows:
\begin{equation}
P^i = \frac{L_B^i p(\bm{r}^i)}{\Sigma_j L_B^j p(\bm{r}^j)/C(\bm{r}^j)}.
\label{eq:pi_prob}
\end{equation}
Here the denominator is a normalization factor,
representing the sum of the weighted probabilities over all galaxies in the 3D localization volume.
As described in subsection~\ref{sec:obs}, we selected our targeted galaxies based on the GLADE catalog.
The GLADE catalog is not necessarily complete. In other words, even if we could observe all the GLADE galaxies,
the survey completeness would still be less than 100\% because of the incompleteness of the catalog.
To take this effect into account, we applied a correction for the catalog incompleteness by introducing the completeness function $C(\bm{r})$.
This function represents the ratio between the sum of the $B$-band luminosities of cataloged galaxies and the total expected $B$-band luminosity within a radial shell at a distance $\bm{r}$ (see also \cite{Sasada:2021tn}).

The weighted probability for each targeted galaxy is given in appendix~\ref{sec:appendix_gallist}. 
By summing the weighted probabilities of the observed $\numofmoircsobsgalaxies$ galaxies, the total completeness of our survey for \gw\ was estimated to be \completeness\%.

\subsection{Candidate screening}
\label{sec:screening}
To find EM counterpart candidates in the reduced MOIRCS data, we conducted two types of analyses:
(1) visual inspection done almost in real-time during the observing run and (2) catalog matching done afterwards.
Both processes require archival data taken previously. 
The archival data used in these analyses include PS1 and DECaPS2 in the $Y$ band, as well as VHS, VVVX, and 2MASS in the {\Ks} band.

For the real-time visual inspection, we used our dedicated image server \citep{Sasada:2021tn}, which visualizes the archival and our new images.
The image server also has a function of image subtraction with the HOTPANTS software \citep{hotpants2015ascl.soft04004B}.
However, due to the lack of the deep reference $Y-$band and {\Ks}-band images over our survey area, we used the image subtraction method only for offline analyses if needed. 
Thus, our real-time transient detection was based on the visual inspection.
The image server generated new (MOIRCS) and reference cutout images around the targeted galaxies. 
By comparing these images, visual inspection was performed to select candidate transients around about 1 arcmin from the GLADE galaxies.

For the catalog match, we performed object matching and brightness comparison between the previous archival and our MOIRCS data.
We searched for possible counterparts by selecting two types of sources depending on the brightness in the MOIRCS data and the depths of the archival data. 
The first type consists of sources detected in both the MOIRCS and archival data.
Although these are not necessarily transients in the strict sense, as they are present in both datasets,
we consider them as candidate transients if
(a) they are detected with high significance in the archival images at the MOIRCS source position, and
(b) the brightness difference between the MOIRCS and archival data exceeds $\thresigmaforcounterpart\sigma$,
suggesting a statistically significant flux variation.
The second type consists of sources that are clearly detected in the MOIRCS data but not in the archival data.
For these, we consider them as candidate transients if their MOIRCS magnitude is brighter than the $\sigmaforvarithre\sigma$ limiting magnitude of the archival data.
Without this criterion, a large number of static faint sources undetected in the shallow archival data would contaminate the candidate transients list.

Several potential candidates were initially identified through the visual inspection process. 
These were further examined to remove likely contaminants, including moving objects or objects misidentified due to differences in filter characteristics between the reference and observed images.
We also excluded a small number of candidates that exhibited features consistent with detector artifacts, such as hot pixels or residual noise.

On the other hand, the catalog-based search yielded no compelling candidates.
The lack of candidates in the catalog-based search is primarily due to the relatively shallow limiting magnitudes of the archival data compared to the MOIRCS observations, 
which made it difficult to satisfy the variability criteria described above.

\section{Results}
\label{sec:results}
\begin{figure*}[!ht]
  \begin{center}
     \includegraphics[width=2\columnwidth]{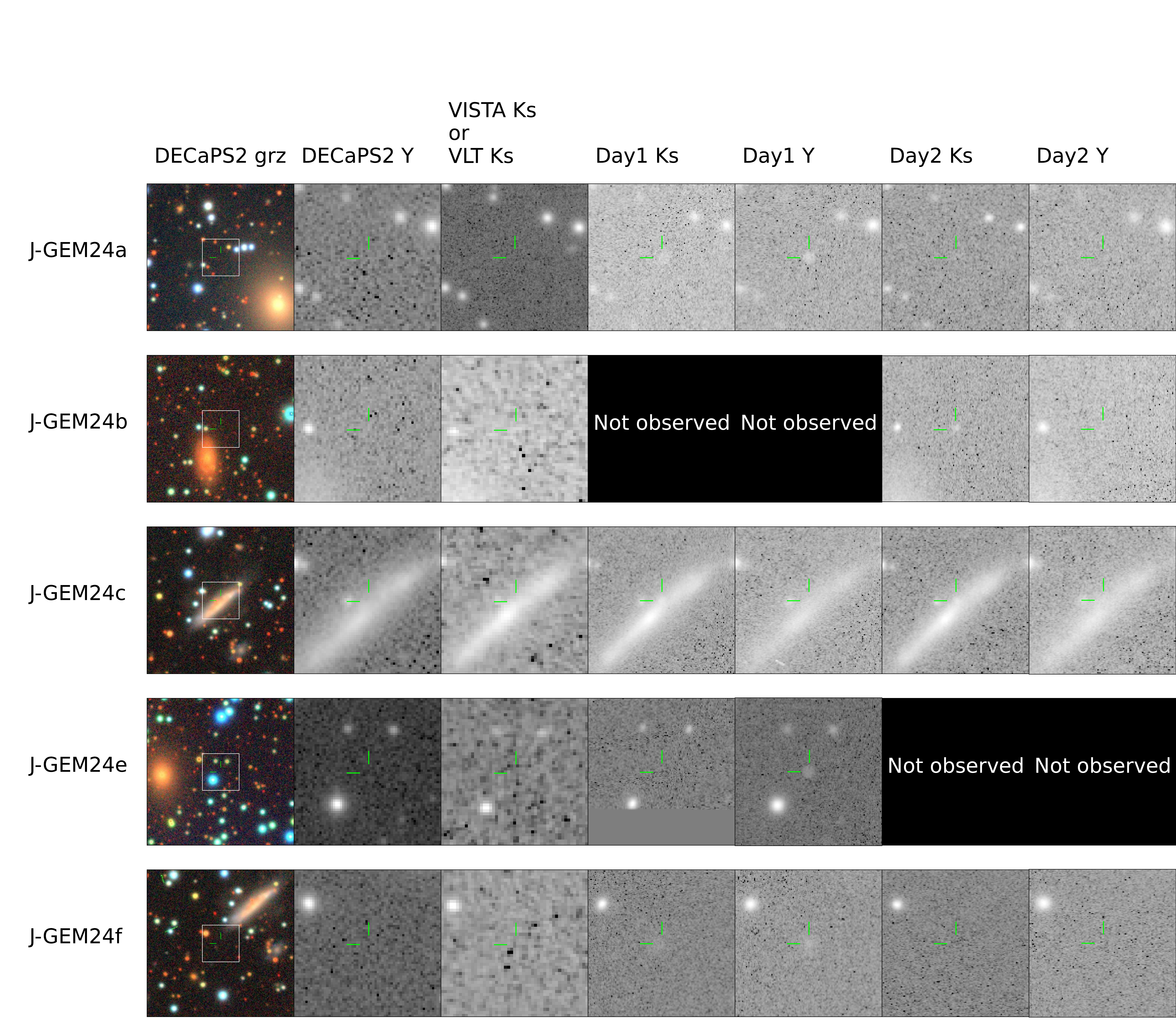}
  \end{center}
  \caption{%
  Cutout images of the EM counterpart candidates identified in our follow-up observations, including reference images (columns 1--3). Each row corresponds to one candidate (from top to bottom: J-GEM24a, 24b, 24c, 24e, and 24f), while each column (from left to right) shows: (1) a DECaPS2 color-composite image created from $g$, $r$, and $z$ bands; (2) a $Y$-band reference image from DECaPS2; (3) a {\Ks}-band reference image taken from either VISTA or VLT archival data; and (4–7) MOIRCS follow-up images taken on two nights in the order of Day 1 {\Ks}, Day 1 $Y$, Day 2 {\Ks}, and Day 2 $Y$. The white square in the DECaPS2 $grz$-band panel indicates the $16.8\arcsec \times 16.8\arcsec$ square field, corresponding to the cutout images to its right. For J-GEM24a, the {\Ks}-band reference image is substituted with a follow-up image obtained by VLT six days after our observations.
  {Alt text: Cutout images of five electromagnetic counterpart candidates. Rows represent individual candidates. Columns show archival images in optical and near-infrared bands, followed by MOIRCS follow-up images from two nights.}
  }%
  \label{fig:images}
\end{figure*}

\begin{figure}[thbp]
  \begin{center}
     \includegraphics[width=\columnwidth]{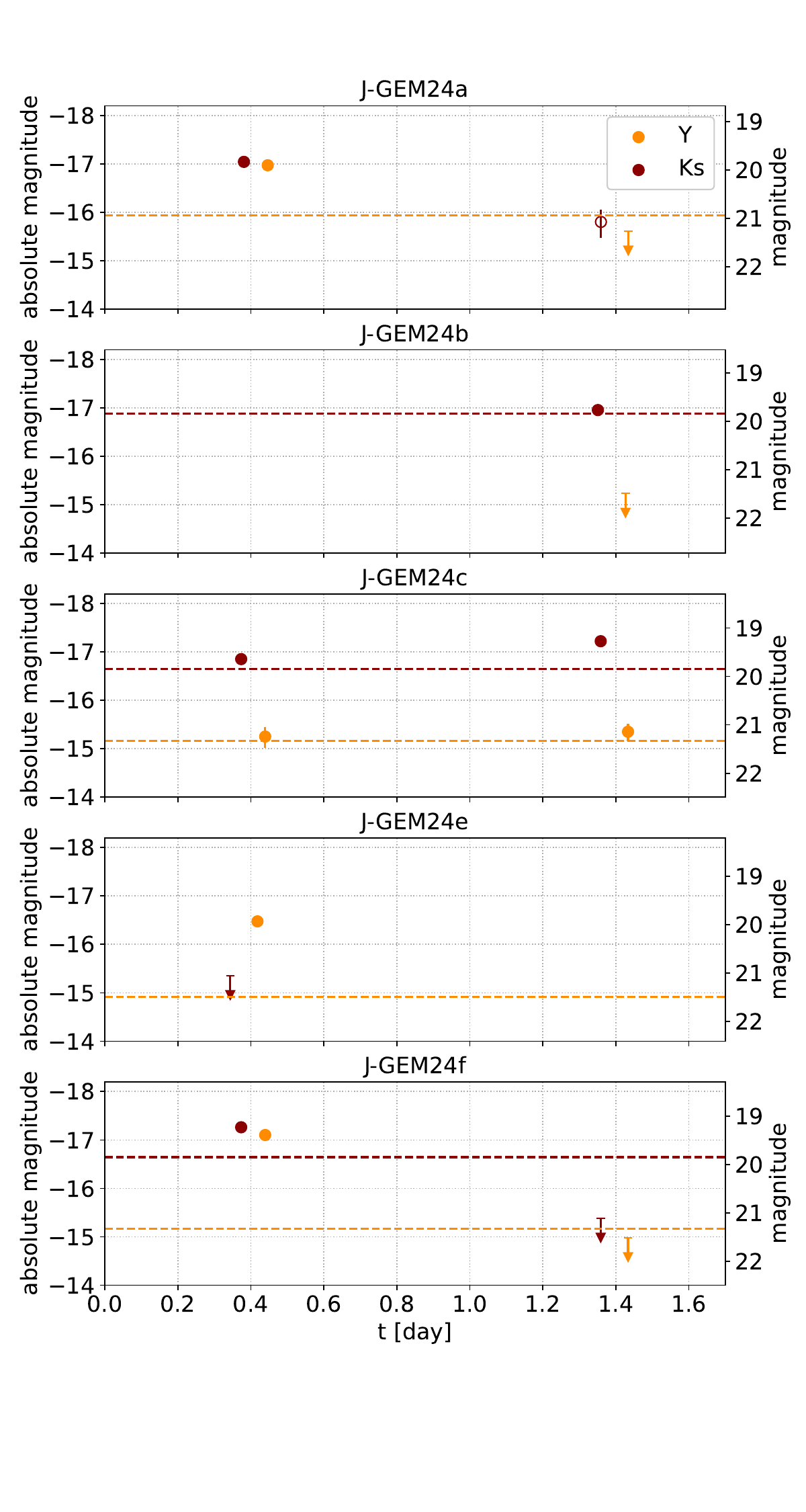}
  \end{center}
  \caption{%
    Photometric measurements of the EM counterpart candidates (J-GEM24a, 24b, 24c, 24e, and 24f) from our follow-up observations. The orange and dark red points represent detections in the $Y$ and {\Ks} bands, respectively, while downward arrows indicate 3$\sigma$ upper limits for non-detection. An open (white-filled) marker indicates a marginal detection with S/N between 3 and 5. Horizontal dashed lines show the 3$\sigma$ limiting magnitudes of the reference images in each band.
    {Alt text: Photometric measurements of five candidate objects, each shown in a separate panel. Magnitudes in Y band and K sub s band are plotted for two nights. Symbols indicate detections, a marginal detection, and upper limits.}
    }
  \label{fig:abs_mags}
\end{figure}

\begin{figure}[thbp]
  \begin{center}
    \includegraphics[width=\columnwidth]{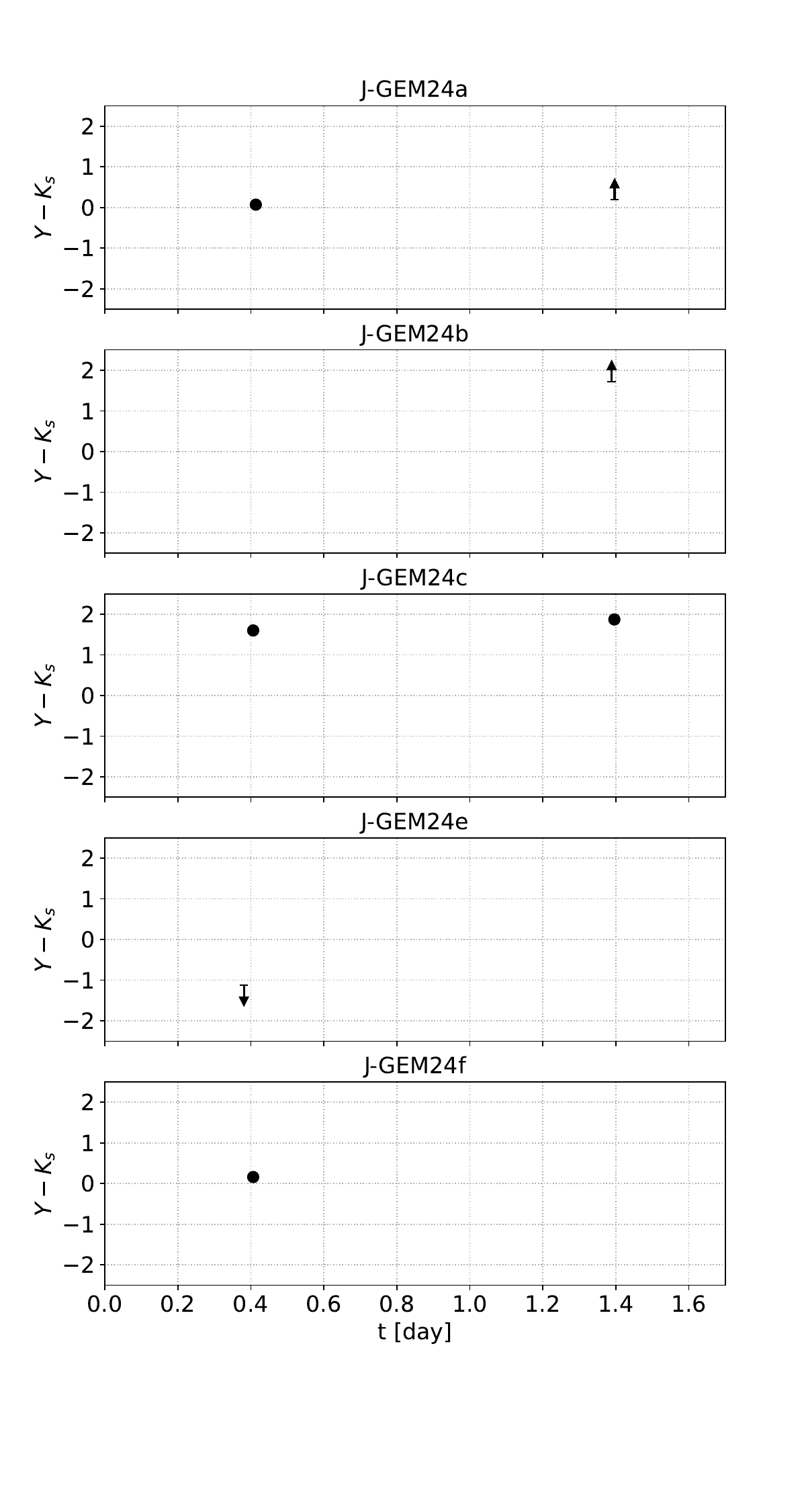}
  \end{center}
  \caption{%
    Color evolution of the EM counterpart candidates (J-GEM24a, 24b, 24c, 24e, and 24f from top to bottom), shown as $Y - K_{\rm s}$ versus time since the GW event.  
    Filled circles indicate color measurements, while arrows represent color limits due to non-detections in either band.
    {Alt text: Color evolution of five candidate objects shown in separate panels. Each panel plots Y minus K sub s color over time. Symbols indicate measured colors or limits due to non-detections.}
    }
  \label{fig:colors}
\end{figure}

As a result of the transient search, we find five candidate transients.
These five EM counterpart candidates are denoted as J-GEM24a, 24b, 24c, 24e, and 24f throughout this paper (J-GEM24d was initially selected as a candidate, but later turned out to be not significant).
All of them are found by real-time visual inspection, while no candidates are found by catalog matching due to the relatively shallow limiting magnitude of the archival catalogs.
Cutout images of the candidates are shown in figure~\ref{fig:images}.
Detailed information about the candidates is summarized in table~\ref{tab:candidates_coordinate} and photometric information is summarized in table~\ref{tab:candidates_photo}.
Figures \ref{fig:abs_mags} and \ref{fig:colors} show the absolute magnitudes and $Y-K_{\rm s}$ color of the detected candidates, respectively. The absolute magnitudes are calculated under the assumption that each candidate is associated with its targeted host galaxy.
No correction for Galactic extinction has been applied to the photometric data: the Galactic extinction is estimated to be $E(B-V) = 0.2$--0.6 toward the direction of the transients, corresponding to $A_Y =$ 0.2--0.6 mag and $A_{K_{\rm s}} =$ 0.06--0.18 mag.

Among the five objects, J-GEM24a and J-GEM24f are detected in both $Y$ and {\Ks} bands on Day 1.
On Day 2, however, these objects are not detected with high statistical significance.
J-GEM24c is detected near the center of the candidate host galaxy and its detection is confirmed by image subtraction both on Day 1 and Day 2 in $Y$ and {\Ks} bands.

The other two candidates (J-GEM24b and 24e) are each detected only once, namely in either the $Y$ or {\Ks} band and on either Day 1 or Day 2.
In fact, J-GEM24b is selected as a candidate in {\Ks} band (19.77 mag), but the VISTA reference image is not deep enough to detect a source of this brightness.
Therefore, J-GEM24b can be a red star that is detected only in our deep MOIRCS data. Thus, we do not include this object for further discussion.

We also measure the shape of the detected objects.
The shape of J-GEM24f appears marginally extended compared to the PSF of the surrounding stars.
However, its FWHM is consistent within 2$\sigma$ of the mean PSF size derived from surrounding point sources (see appendix~\ref{sec:appendix_measurements} for details).
The shapes of the other candidates are fully consistent with a point source.

For the three objects with multiple detections (J-GEM24a, 24c, and 24f), we also measure their positions and motions on the sky to test whether they can be moving objects.
As a result, we do not find any significant ($> 3 \sigma$) motions for these candidates as compared with the surrounding stars (see appendix~\ref{sec:appendix_measurements} for the details).

Major contamination in transient surveys is moving objects. 
Thus, we first cross-match our candidates against asteroids with known orbits using the MPChecker\footnote{MPChecker $\langle$\url{https://minorplanetcenter.org/cgi-bin/checkmp.cgi}$\rangle$.}, but our candidates do not match any known asteroids.
However, all of our candidates have about 20 AB mag  in $Y$ or {\Ks} bands, and the database of asteroids is not necessarily complete for such faint objects.
In the following section, we discuss their possible natures including extragalactic and Galactic transients.

%
\begin{table*}[!ht]
\tbl{Basic properties of the EM counterpart candidates identified in our follow-up observations.}{
\begin{tabular}{lllllllll}
\hline
\multicolumn{3}{l}{candidate} & \multicolumn{4}{l}{host galaxy candidate} & \multicolumn{2}{l}{separation} \\
\hline
name & RA (J2000.0) [$^\circ$] & Dec (J2000.0) [$^\circ$] & galaxy name & RA (J2000.0) [$^\circ$] & Dec (J2000.0) [$^\circ$] & $d_L$\footnotemark[$*$] [Mpc] & [\arcsec] & [kpc] \\
\hline
J-GEM24a & 122.215845 & $-$24.516397 & GL080850$-$243120 & 122.207771 & $-$24.522257 & 237 & 33.8 & 38.9 \\
J-GEM24b & 120.153077 & $-$26.334192 & GL080037$-$262017 & 120.154976 & $-$26.337975 & 221 & 14.9 & 16.0 \\
J-GEM24c & 123.297671 & $-$28.042531 & GL081312$-$280235 & 123.297997 & $-$28.042965 & 199 & 1.9 & 1.8 \\
J-GEM24e & 122.098393 & $-$31.049748 & GL080826$-$310301 & 122.106964 & $-$31.050245 & 191 & 26.5 & 24.5 \\
J-GEM24f & 123.302842 & $-$28.048007 & GL081312$-$280235 & 123.297997 & $-$28.042965 & 199 & 23.8 & 22.9 \\
\hline
\end{tabular}
}\label{tab:candidates_coordinate}
\begin{tabnote}
\footnotemark[$*$] Luminosity distance of the host galaxy candidate, taken from the GLADE catalog.
\end{tabnote}
\end{table*}

\begin{table*}[ht]
\tbl{Photometric data of the EM counterpart candidates in our MOIRCS observations.}{
\begin{tabular}{llllllllll}
\hline
candidate & obs. date (UT) & MJD & filter & RA (J2000.0) [$^\circ$] & Dec (J2000.0) [$^\circ$] & mag & mag error & $S/N$ & limiting mag (3$\sigma$)\\
\hline
J-GEM24a & 240423 06:44:04.5 & 60423.28061 & {\Ks} & 122.215786 & $-$24.516459 & 19.83 & 0.13 & 8.7 & 20.98 \\
         & 240423 08:18:00.1 & 60423.34583 & $Y$ & 122.215845 & $-$24.516397 & 19.90 & 0.10 & 10.7 & 21.28 \\
         & 240424 06:12:16.5 & 60424.25852 & {\Ks} & 122.215753 & $-$24.516337 & 21.07 & 0.28 & 3.8 & 21.33 \\
         & 240424 08:00:03.1 & 60424.33337 & $Y$ & --- & --- & --- & --- & --- & 21.26 \\
J-GEM24b & 240424 06:00:15.5 & 60424.25018 & {\Ks} & 120.153077 & $-$26.334192 & 19.77 & 0.11 & 9.7 & 21.04 \\
         & 240424 07:49:08.1 & 60424.32579 & $Y$ & --- & --- & --- & --- & --- & 21.49 \\
J-GEM24c & 240423 06:33:13.5 & 60423.27307 & {\Ks} & 123.297678 & $-$28.042520 & 19.64 & 0.08 & 14.0 & 21.07 \\
         & 240423 08:07:59.6 & 60423.33888 & $Y$ & 123.297578 & $-$28.042592 & 21.24 & 0.22 & 5.0 & 21.38 \\
        & 240424 06:11:21.5 & 60424.25789 & {\Ks} & 123.297671 & $-$28.042531 & 19.27 & 0.05 & 23.5 & 21.11 \\
        & 240424 07:59:12.1 & 60424.33278 & $Y$ & 123.297700 & $-$28.042470 & 21.14 & 0.17 & 6.3 & 21.51 \\
J-GEM24e & 240423 05:50:11.5 & 60423.24319 & {\Ks} & --- & --- & --- & --- & --- & 21.05 \\
        & 240423 07:37:33.6 & 60423.31775 & $Y$ & 122.098393 & $-$31.049748 & 19.93 & 0.10 & 10.4 & 21.33 \\
J-GEM24f & 240423 06:33:13.5 & 60423.27307 & {\Ks} & 123.302958 & $-$28.048073 & 19.23 & 0.07 & 15.9 & 21.07 \\
        & 240423 08:07:59.6 & 60423.33888 & $Y$ & 123.302842 & $-$28.048007 & 19.39 & 0.06 & 18.7 & 21.38 \\
        & 240424 06:11:21.5 & 60424.25789 & {\Ks} & --- & --- & --- & --- & --- & 21.11 \\
        & 240424 07:59:12.1 & 60424.33278 & $Y$ & --- & --- & --- & --- & --- & 21.51 \\
\hline
\end{tabular}
}\label{tab:candidates_photo}
\end{table*}
%

\section{Discussion}
\label{sec:discussion}
\subsection{Nature of the EM counterpart candidates}
\subsubsection{Extragalactic transients}
\label{sec:extragalactic}
\begin{figure*}[thbp]
  \begin{center}
     \includegraphics[width=2\columnwidth]{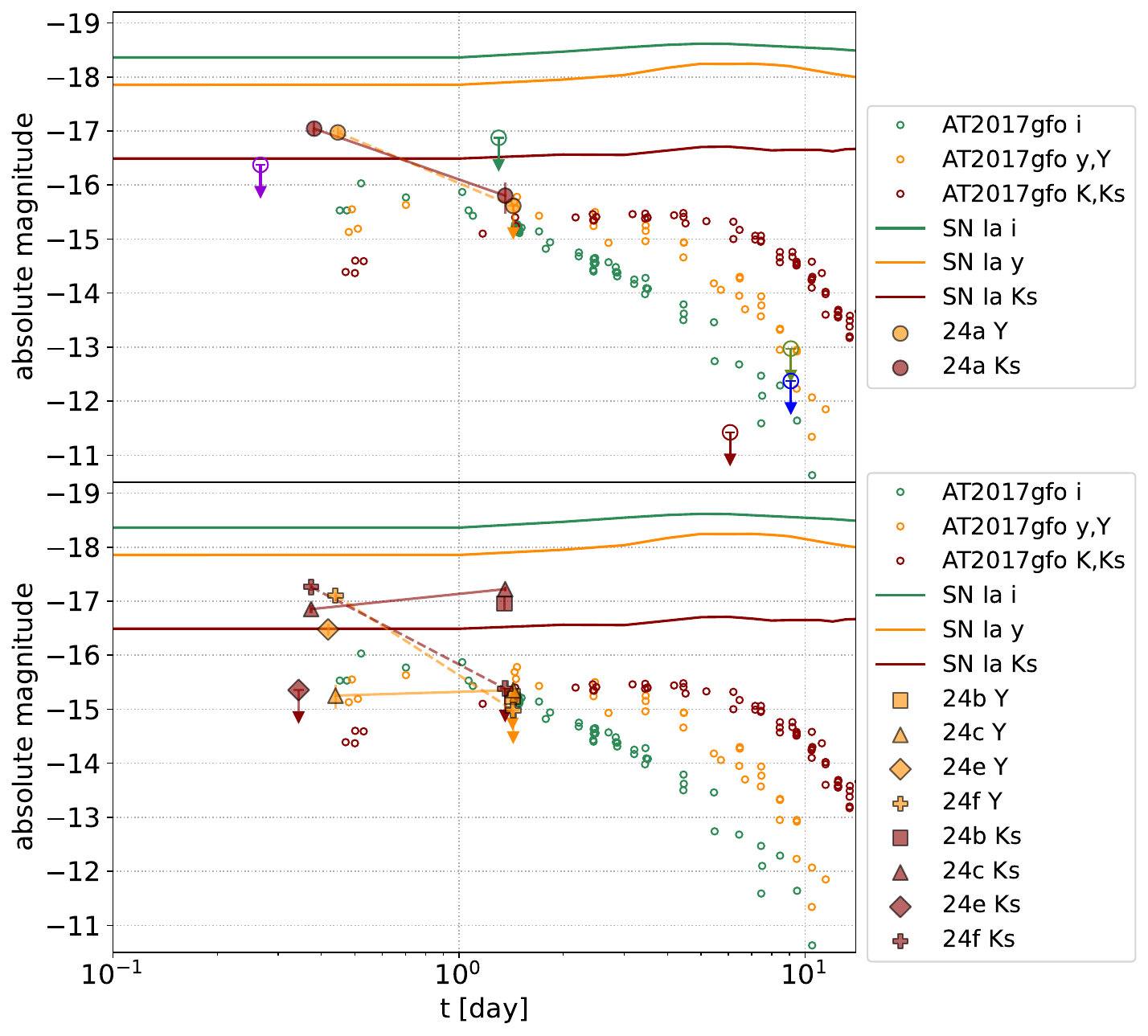}
  \end{center}
  \caption{%
    Comparison of the absolute magnitudes of the EM counterpart candidates with those of kilonova AT2017gfo and a typical Type Ia SN. The upper panel shows J-GEM24a, while the lower panel displays the other candidates (J-GEM24b, 24c, 24e, and 24f). Orange and dark red symbols represent measurements in the $Y$ and {\Ks} bands, respectively. 
    Solid lines connect detections, while dashed lines indicate connections involving upper limits.
    Open circles in the upper panel denote upper limits from Gemini and VLT reported via GCN Circulars; the dark red circle corresponds to the VLT {\Ks}-band observation, while the other four points are from Gemini, shown in purple ($g$ band), blue ($r$ band), green ($i$ band), and olive ($z$ band) respectively.
    {Alt text: Two-panel figure comparing absolute magnitudes of five candidate objects with those of a kilonova and a Type Ia supernova. The upper panel shows one candidate, and the lower panel shows four others. Time since the gravitational wave event is on the horizontal axis. Symbols indicate detections or upper limits in multiple photometric bands, including our follow-up observations and archival constraints reported from other telescopes.}
  }
  \label{fig:abs_mag_range}
\end{figure*}

We begin by considering the extragalactic nature of the EM counterpart candidates.
Figure~\ref{fig:abs_mag_range} compares the absolute-magnitude light curves of our candidates with those of typical Type Ia supernovae (SNe) and the kilonova (AT2017gfo, \cite{abbott17MMA}).
Type Ia SN light curves are calculated with the SALT2 template (\cite{salt2_2007}) through the {\tt sncosmo} package (\cite{sncosmo2016}).
As shown in the figure, the absolute magnitudes of our candidates are in the range of $-15$ to $-17$ mag.
If they are extragalactic transients, their brightness is broadly consistent with the peak of kilonovae or fainter part of SNe.

Below we give more detailed discussion on J-GEM24a, 24c, and 24f with multiple detections.
In figure~\ref{fig:abs_mag_range}, both J-GEM24a and 24f show very rapid decline in {\Ks} band. 
For J-GEM24a, optical upper limits are also obtained by Gemini-South/GMOS \citep{ahumada24}.
In addition, a deep NIR image was taken with VLT/HAWK-I 6 days after our observations (PI: ENGRAVE collaboration) and is available in the archive. We do not detect J-GEM24a, giving an upper limit of about $25$ mag in the {\Ks}-band.
Combined with these data, the declining rates of both J-GEM24a and 24f are much faster than that of AT2017gfo.
Thus, we conclude that these two objects are unlikely to be kilonova.

In fact, kilonova is one of the most rapidly evolving extragalactic transients. 
Other rapidly evolving transients include so-called fast blue optical transients (FBOT), but they are more luminous (and less rapid). 
While afterglow of the gamma-ray bursts show faster time evolution than kilonova, no gamma-ray burst event has been reported in the corresponding regions.
Thus, these classes of transients are not consistent with the properties of J-GEM24a and 24f.
Indeed, both of these candidates are located relatively far from the possible host galaxies ($>$ 20--30 kpc, table~\ref{tab:candidates_coordinate}), and the association to the host galaxies is less certain.
Thus, J-GEM24a and 24f may not be extragalactic transients (see \mbox{sub-subsection~\ref{sec:galactic}}).

J-GEM24c shows flat light curves in both the $Y$ and {\Ks} bands, and its color does not evolve with time.
The observed magnitude is significantly brighter than expected brightness of a kilonova. Therefore, J-GEM24c is also not consistent with a kilonova.

In contrast to J-GEM24a and 24f, 
J-GEM24c is located near the center of a host galaxy candidate, making its association with the host galaxy more robust.
The slow evolution of the light curve is quite consistent with normal SNe rather than kilonovae.
A striking difference with SNe is its very red color.
The $Y-K_{\rm s}$ color is about 2 mag, which corresponds to the blackbody temperature of about 1300~K.
This is significantly lower than the expected temperature of SNe.
The Galactic reddening along the line of sight is $E(B-V)_{\rm MW} = 0.4$~mag based on the SFD98 map \citep{Schlegel1998}. 
Even after accounting for this foreground extinction, the object still shows an unusually red color.
To recover the typical color of normal SNe, $J-K_{\rm s} \sim 0$ (see \cite{SNIa_NIR2019}), additional extinction in the host galaxy of $E(B-V)_{\rm host} \sim 2.7$ is required.
This implies that the intrinsic absolute magnitude is about $-21$ mag, which is comparable to that of superluminous SNe.
Therefore, although J-GEM24c seems to be an extragalactic transient, its nature is not entirely clear.

Finally, we comment on the expected event rate of SNe in our galaxy targeted survey.
The number of candidate host galaxies observed is approximately 200. 
Assuming a typical SN rate of 1--2 events per century per galaxy (e.g., \cite{adams13}),
the expected number of SNe across all these galaxies is $\sim$2--4 per year. By further assuming that a SN remains detectable for about two months, the expected number of such events during our two-day observing window is approximately (2--4) yr$^{-1}$ × (2/12) yr $\sim$ 0.3--0.6 SNe.
Thus, it is not very surprising to find one SN in our survey. 
However, as discussed above, the properties of J-GEM24c may not be consistent with normal SNe.

The expected small number of extragalactic SNe also implies that it is unlikely that all of our five candidates are extragalactic SNe. Thus, we discuss the possibility of Galactic transients in the next section.

%

\subsubsection{Galactic transients}
\label{sec:galactic}
In this section, we consider the possibility that the candidate transients originate from Galactic sources.
In fact, our survey area is largely overlapped with the Galactic plane, and Galactic transients such as stellar flares, classical novae, or dwarf novae (DNe) may appear in our survey fields.
Among these, stellar flares are less likely to be detected as transients due to their typically small amplitude ($\Delta K_{\rm s}$ < 0.1 mag) in the NIR band (\cite{Davenport2012}). 

Classical novae are considered as potential contaminants in our Galactic-plane fields.
They typically exhibit relatively slow decline timescales, with $t_3$ values (time to fade by 3~mag from peak) ranging from 10 to 100 days in optical \citep{warner95}.
Assuming that the NIR light curves follow similar fading timescales, we adopt a visibility window of 10--100 days for classical novae within the magnitude range of our candidates ($K_{\rm s} \sim 18$--21 AB mag).
\citet{De2021} derived a Galactic nova rate of $43.7^{+19.5}_{-8.7}$ yr$^{-1}$ based on wide-field NIR observations.
Assuming that these novae are uniformly distributed within the Galactic plane ($|b| < 10^\circ$, corresponding to $\sim 7200$ deg$^2$), this translates to a surface rate of $\sim 1.7 \times 10^{-5}$ deg$^{-2}$ day$^{-1}$.
Given our effective survey area of $\sim 0.06$ deg$^2$ (200 galaxies $\times$ 1 arcmin$^2$ per galaxy), and adopting a visibility window of 10--100 days, the expected number of classical novae in our survey is $\lesssim 10^{-4}$.
We therefore consider the likelihood of contamination from classical novae to be negligible.
Furthermore, the relatively slow fading of classical novae makes them inconsistent with the observed rapid decline of our candidates, suggesting they are unlikely contaminants.

\begin{figure}[htbp]
  \begin{center}
     \includegraphics[width=\columnwidth]{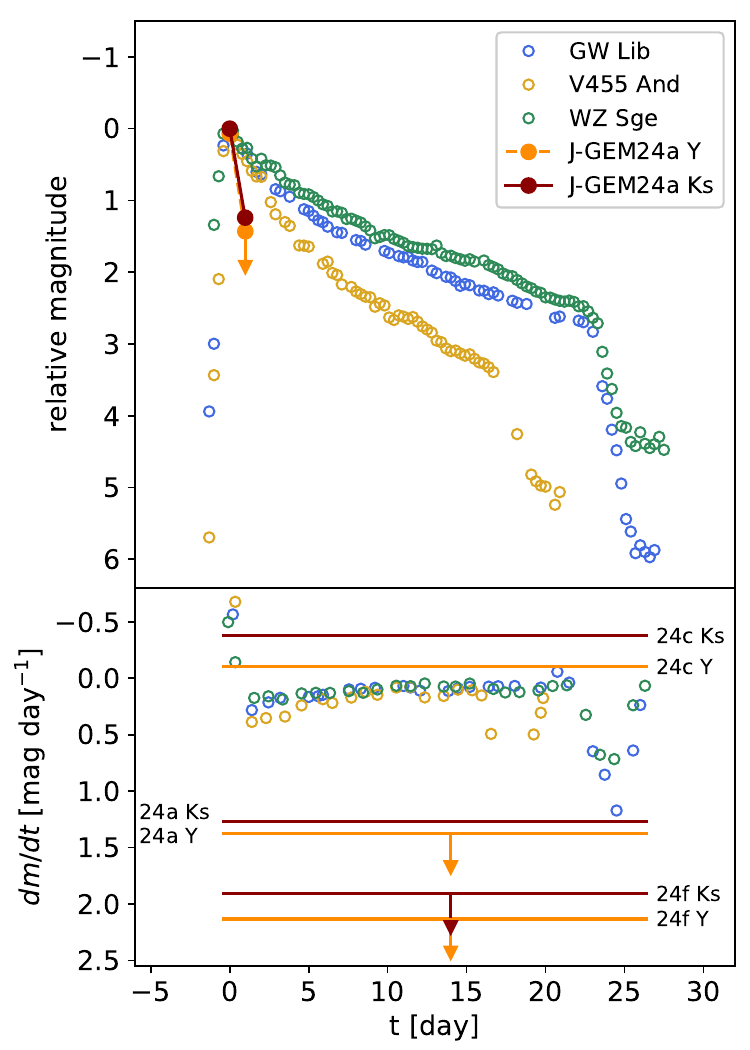}
  \end{center}
  \caption{%
    Comparison between the light curve slopes of three well-observed WZ Sge-type DNe and those of our candidates. The upper panel shows smoothed light curves of GW Lib , V455 And, and WZ Sge, \citep{kato09} created from various photometric data (e.g., $V$-band, unfiltered) with vertical offsets to match the $V$-band data, and further adjusted so that the peak magnitude is set to zero. The light curves of J-GEM24a in the $Y$ and {\Ks} bands are also shown for comparison. The lower panel shows the light curve slope ($\mathrm{mag\,day^{-1}}$). Horizontal lines represent the slopes derived from the observed magnitudes of each candidate in the $Y$ and {\Ks} bands. Horizontal lines with downward arrows indicate lower limits on the slope, derived from two-epoch measurements where one of the magnitudes is an upper limit.
    {Alt text: Two-panel figure comparing light curves and light curve slopes of three known dwarf novae with those of observed candidates. The upper panel includes normalized light curves of dwarf novae and the added light curves of J-GEM24a.
    In the lower panel, horizontal lines are overlaid to represent slopes for each candidate. Lines with downward arrows indicate lower limits based on two-epoch photometry with non-detections.}
    }
  \label{fig:DN_LCs}
\end{figure}

DNe are expected to be the most common transient objects near the Galactic plane. As high-amplitude transients, we compare light curves of DN superoutbursts with the three candidates that have multiple observations (J-GEM24a, 24c, and 24f).
Figure~\ref{fig:DN_LCs} shows example superoutburst light curves (top) and rising/declining rate (bottom), mainly based on $V$-band data.   
Although the decline amplitude in the NIR is generally smaller than in optical due to disk cooling and reddening during the outburst, the overall temporal behavior remains similar.  
Indeed, \citet{Matsui2009PASJ...61.1081M} show that the $V$- and $J$-band light curves of the WZ Sge-type DN V455 And follow a similar trend, with a $\sim$1 mag difference in fading amplitude.  
Thus, our comparison using the $Y$ and {\Ks} bands is conservative.

J-GEM24a and 24f show a rapid decline ($dm/dt > 1$ in both $Y$ and {\Ks} bands) and the declining rate is significantly faster than that of DN superoutburst light curves \citep{Patterson97,Kato2002,Otulakowska-Hypka2016}.
Thus, these two candidates are not likely to be DN superoutbursts.

Note that, more frequent, normal outbursts of DNe can show a faster decline depending on the phase.
In fact, the properties of J-GEM24f on Day~1 and 2 (declining rate and color) can be similar to those of normal outbursts.
For J-GEM24a, on the other hand, the deep NIR upper limit obtained with VLT/HAWK-I (see \mbox{sub-subsection~\ref{sec:extragalactic}}) is more than 5 magnitudes fainter than its brightness on Day~1. Such a large decline in magnitude over just 6 days is difficult to reconcile even with DN normal outbursts.

J-GEM24c shows a moderate rise in the light curve, which is consistent with the DN superoutbursts just before the peak. 
Note that J-GEM24c was discovered just on top of the galaxy (GL081312$-$280235) with about 2 arcsec separation from the center. 
The probability of finding Galactic transients on top of the external galaxy is rather low, i.e., an area of the central part of the galaxy is only 0.1\%--1\% in the 1 arcmin search region around the galaxy.
Thus, it may be more likely that J-GEM24c is an extragalactic, red transient as discussed in \mbox{sub-subsection~\ref{sec:extragalactic}}.
The other two candidates (J-GEM24b and 24e) are detected only once, and it is difficult to evaluate the nature based on the light curve.

To further examine the possibility of DNe, we here estimate an expected number of DN superoutbursts by following \citet{rau07}.
Note that an expected event rate of normal outburst is even higher than this estimate, and thus, the following estimate gives a lower limit of the expected number of DN outbursts in general.

In the quiescent phase, DN systems have an absolute magnitude of $M_R \sim 12$ mag.
The DN superoutbursts show the amplitude of about 6 mag, i.e., the peak absolute magnitude of $M_R \sim +6$ mag. 
The peak magnitude in NIR is similar ($M_Y \sim +$ 6 mag) due to the blue color of the superoutburst ($R-Y \sim 0$, as expected from $T \sim 10^4$ K, \cite{warner95}).
Thus, the superoutburst would be detectable with our search (21 mag depth in $Y$-band) 
if the distance to the object is $< 10$ kpc.
Also, the object in the quiescent phase
($M_Y \sim 11$ mag, by assuming a red color, $R-Y \sim 1$ mag)
would be visible in the reference images
if the distance to the object is $< 1$ kpc.
Thus, if the source is located between 1 kpc and 10 kpc, 
the superoutburst may look like a transient in our MOIRCS images without any counterpart source  in the reference images.

To estimate the expected event rate, 
we first estimate the survey volume of our observations.
Our survey orientation is centered on the Galactic coordinate of $l \simeq 250^{\circ}$  and $b \simeq +5^{\circ}$.
Thus, if we assume a height of the Galactic disk to be about 300 pc and no DN system is located beyond this height, the maximum distance to detect the DN would be about 3.5 kpc.
Assuming $1 \times 1 \ {\rm arcmin^2}$ region around the targeted galaxy (a typical separation from the galaxy is about 30 arcsec, see table~\ref{tab:candidates_coordinate}), 
the observed volume between 1 and 3.5 kpc is about $10^3 \ {\rm pc^3}$. 
Thus, the total effective volume of our survey for DN systems is $V \sim 2 \times 10^5 \ {\rm pc^3}$.

Then, we estimate the number of DN superoutbursts within this volume.
Observational and theoretical studies loosely constrain the local number density of the DN system to be in the range of $\rho =$ (0.03--1) $\times 10^{-3} \ {\rm pc^{-3}}$  \citep{kolb93,dekool93,schwope02}.
Thus, the number of DN systems in our survey volume is $N_{\rm DN} \sim$ 6--200.
Assuming that the duty cycle of the superoutburst to be 1 yr, and the outburst is visible over 10 days (as adopted by \cite{rau07}), the number of observed DN superoutbursts for a certain time is $N_{\rm outburst} \sim$ 0.2--6
\footnote{Considering the Galactic longitude of our survey direction ($l \sim 250$ deg), the Galactocentric distance of the Galactic position at 3.5 kpc away is about $R = 10$ kpc. 
Thus, as compared with the Sun's location (Galactocentric distance of $R = 8.5$ kpc), the local stellar density may be somewhat smaller, but it is still within a factor of about 2 assuming the scale length of the Galactic disk is $R_d = 2.9$ kpc (i.e., $\rho \propto \exp(-R/R_d)$).}.
As described above, the frequency of normal outbursts is even higher than this.
Therefore, although the event rate of DN system and their outburst frequency are uncertain, it is not surprising to detect a few DN outbursts in our observations.

%

\subsubsection{Other possibilities}
\label{sec:moving}
In the final part of this discussion, we also consider alternative possibilities that the observed candidates are caused by moving objects, such as solar system bodies or artificial objects in Earth orbit, including satellites and debris.

As described in subsection~\ref{sec:screening}, 
some of the candidates detected more than once show marginal indications of motion.
In fact, for both J-GEM24a and 24f, the positions in the $Y$- and {\Ks}-band images on Day 1 show marginal motion between the two observations, at the $\sim$2--3$\sigma$ level relative to the positional scatter of nearby reference stars. For J-GEM24a, a similarly marginal displacement is also seen between Day 1 and Day 2, suggesting no clear motion. In contrast, for J-GEM24f, if the marginal shift on Day 1 reflects real motion of $\sim$6.7 arcsec/day, the object should appear at a different position on Day 2, but no such detection is made.

Note that the Ecliptic latitudes of all of our candidates range from $-50^\circ$ to $-42^\circ$:
they are away from the Ecliptic plane, where most asteroids exist.
Therefore, while it is unlikely that our candidates are typical asteroids or common Solar System objects, the possibility remains that they could be comets or other less frequently observed Solar System objects.

The candidates exhibit $Y-K_{\rm s}$ colors of approximately 0 in the AB magnitude system, corresponding to $Y-K_{\rm s}$ $\sim$ 1.2 in the Vega system. This value is broadly consistent with the color distribution of known asteroids presented in the large-scale NIR catalog MOVIS-C (\cite{MOVIS2016A&A...591A.115P}), and falls within the redder end of the asteroid population. 
However, the absence of significant apparent motion suggests that the candidates are unlikely to be a typical main-belt asteroid (MBA) or near-Earth object (NEO). While a decline of over one magnitude within a day can occur in rare cases, such as in contact binaries or extremely elongated bodies, the occurrence rate of such asteroids is very low (\cite{Duffard2009A&A...505.1283D, Thirouin2014A&A...569A...3T}). Therefore, although not impossible, the candidates are unlikely to be asteroids.

The observed characteristics of these candidates may initially appear consistent with an outbursting long-period comet (LPC), but a more detailed examination suggests that this interpretation is also unlikely. LPCs are known to undergo sudden increase or decrease in brightness due to rapid gas and dust release, typically near perihelion \citep{Sekanina2017arXiv171203197S}. 
However, our candidates appear as point sources, lacking any extended coma.
The extremely slow apparent motion ($<$ 1 arcsec/hour) suggests a heliocentric distance greater than 10 AU.
At such distances, cometary activity is usually driven by the sublimation of supervolatile ices (e.g., CO, CO$_2$), and the resulting photometric changes occur gradually over days to weeks, not typically showing rapid fading within a single day \citep{Jewitt2017ApJ...847L..19J,Hui2019AJ....157..162H}. 
Taken together, the stellar PSF, slow motion, and rapid fading suggest that these candidates are unlikely to be an LPC undergoing an outburst.

Finally, we consider a possibility that the candidate transients originate from specular reflections (glints) caused by artificial satellites or space debris. Glints occur when the Sun, the satellite, and the observer are aligned in a specific geometric configuration, resulting in very brief, highly directional, and transient events. A statistical analysis by \citet{Corbett2020ApJ...903L..27C} show that glints are frequently detected when the angular separation from the Sun is around 95 degrees. Indeed, all of the candidates identified in this study appeared at solar-object separations between 94$^{\circ}$ and 98$^{\circ}$. However, even geostationary satellites exhibit significant motion in the equatorial coordinate system (RA/Dec), making it virtually impossible for a glint to reappear at the same coordinates after an interval of just over one hour. Therefore, we conclude that events J-GEM24a, 24c, and 24f are unlikely to originate from artificial satellites. In contrast, J-GEM24e, which was detected only once, remains a possible glint candidate; however, no matching objects are found in cross-matching with existing satellite and debris catalogs (Space-Track \footnote{Space-Track $\langle$\url{https://www.space-track.org}$\rangle$.}).

To conclude this discussion, we summarize the possible nature of the candidates.
None of the candidates exhibits behavior consistent with a kilonova.
All candidates except for J-GEM24a can be attributed to known types of transients:
J-GEM24b is likely a red star (non-transient), J-GEM24c appears to be consistent with a reddened SN (albeit high luminosity), 
J-GEM24e (only one detection) can be either asteroids or satellite glints, and J-GEM24f could be a DN.
As shown in figure~\ref{fig:abs_mag_range}, J-GEM24a is not detected in the g-band several hours prior to our observations, suggesting that it is a red and rapidly evolving transient.

\subsection{Sensitivity to kilonova}
Although the probability that \gw\ includes a NS turned out to be low, we discuss how our NIR observations could constrain the properties of kilonova by assuming that it was a real event.
Search for kilonova in NIR has several advantages:
(1) Kilonova can be intrinsically very faint in optical wavelength. 
This may be the case in particular when viewed from the equatorial direction.
In such a case, blue emission can be absorbed by lanthanide-rich dynamical ejecta in the equatorial plane, largely suppressing the blue photons, while bright NIR emission is still expected (e.g., \cite{kawaguchi18,wollaeger18,bulla19,kawaguchi20,kitamura25}).
(2) In general, NIR emission of kilonova shows a slower brightness evolution, lasting for more than 10 days as observed in AT2017gfo (e.g., \cite{chornock17,tanvir17,Utsumi2017}).

To demonstrate the constraining power of our NIR observations, we apply a simple, analytic light curve model as introduced by \citet{metzger17} and \citet{villar17}.
Specifically, we follow the formalism by \citet{kitamura25} by parameterizing the mass of the ejecta $M_{\rm ej}$, ejecta velocity $v_{\rm ej}$, and optical opacity $\kappa$.
For simplicity, we adopt one-component assumption and the velocity of the ejecta is fixed to be $v_{\rm ej} = 0.1c$.

\begin{figure}[thbp]
  \begin{center}
      \includegraphics[width=\columnwidth]{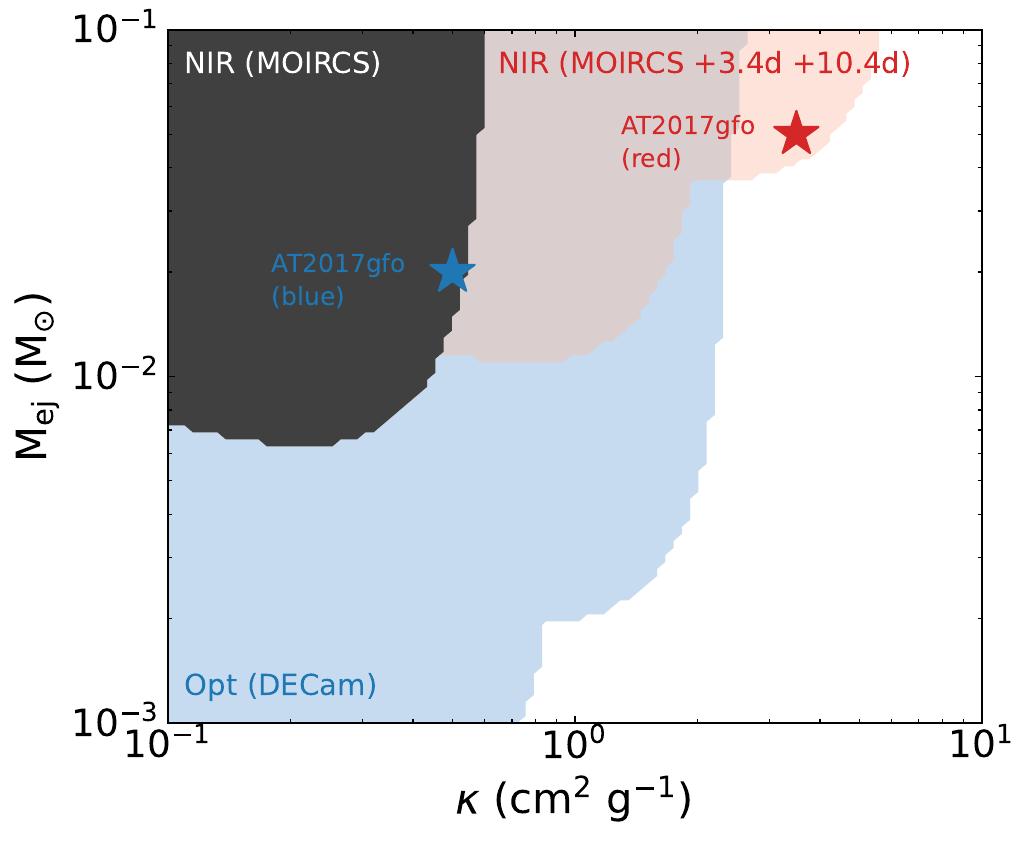}
  \end{center}
  \caption{%
    Expected constraints on the ejecta mass $M_{\rm ej}$ and optical opacity $\kappa$ based on the non-detection of kilonova from our NIR observations and other follow-up observations. The black shaded region shows the constraints obtained from near-infrared non-detections at 0.4 and 1.4 days after the merger, while the red shaded region shows the same from hypothetical near-infrared observations adding 3.4 and 10.4 days after the merger. The blue shaded region shows the constraints from 23 mag optical observations in 6 days after the merger (as done with DECam for \gw, \cite{EMfollowups_Pillas2025}).
    The blue and red symbols show the estimated parameters of "blue" and "red" components for AT2017gfo, respectively \citep{villar17}.
    {Alt text: A plot showing kilonova parameter space with three shaded regions and two colored points, plotted against ejecta mass and optical opacity. The shaded regions indicate constraints from early-time near-infrared non-detections, hypothetical later-time near-infrared observations, and deep optical observations with DECam. Colored points mark the blue and red components of AT2017gfo.}
    }
  \label{fig:param}
\end{figure}

Figure \ref{fig:param} shows the excluded region of the parameters ($M_{\rm ej}$ and $\kappa$) based on the non-detection of kilonova from our NIR ($Y$ and $K$) observations at 0.4 and 1.4 days after the GW event (black shaded region). The distance to the event is assumed to be 200 Mpc.
The region with a high ejecta mass and a low opacity tends to be constrained.
In a simple kilonova model with a low opacity ($\kappa < 1 \ {\rm cm^2 g^{-1}}$), the optical emission tends to be brighter at a few days after the merger while the NIR emission rises slowly.
As a result, $Y$-band limit at 0.4--1.4 days mainly constrains these parameter space.

For \gw, several other follow-up observations have also been performed. \citet{EMfollowups_Pillas2025} presented an extensive summary of their optical and NIR follow-up observations.
In particular, their optical observations with DECam cover $\sim 80\%$ of the localization area to a depth of 23 mag within the first 6 days. The parameter constraint region from their optical observations is shown as the blue shaded region in figure~\ref{fig:param}. Thanks to the higher sensitivity in the optical observations, a parameter space with a small opacity, which corresponds to brighter optical emission, is widely constrained.

Note that \citet{EMfollowups_Pillas2025} also presented wide-field observations with the WINTER telescope, 
which cover 16\% of the localization area to a depth of about 16.5 mag.
Therefore, the constraining power of their NIR observations is more limited compared to ours (which reached about 21 mag).
Also, wide-field follow-up observations have been performed with the PRIME telescope \citep{guiffreda24}. 
They cover about 14\% of the localization area to a depth of about 21 mag in the first two days (see \cite{EMfollowups_Pillas2025}). 
Thus, constraints on the kilonova parameters from their observations would be comparable to ours (black shaded region in figure~\ref{fig:param}).

One of the advantages of the NIR observations is that kilonova emission is expected to last for a long time. To see this effect, we also show the constrained region by assuming that the observations with the same sensitivity were additionally performed at 3.4 and 10.4 days after the merger (red shaded region).
In this case, the constrained parameter space would be extended to a higher opacity region ($\kappa > 1 \ {\rm cm^2 g^{-1}}$).
In such cases, the optical emission tends to be suppressed and the NIR emission becomes dominant.
This demonstrates that deep NIR observations can detect (or constrain) the red kilonova without bright optical emission even at 200 Mpc distance.

%

\subsection{Lessons learned and future prospects}
As discussed in the previous section, NIR follow-up observations have several advantages to search for kilonova.
Our deep NIR observations may provide useful lessons for future NIR searches for GW sources with wide-field ground telescopes, such as WINTER (\cite{WINTER2020}) and PRIME (\cite{PRIME2023}), as well as space telescopes such as Euclid (\cite{euclid}) and Roman (\cite{Roman2019arXiv}).
Here we discuss several challenges and possible improvements based on our follow-up observations.

The first challenge is the lack of deep reference images and catalogs in NIR wavelengths.
As discussed in section~\ref{sec:observations}, we performed transient search by visual check (using the reference images) and by catalog matching (using the catalogs).
However, in both cases, the existing reference data (VISTA) are not as deep as our MOIRCS data in particular in {\Ks} band.
Thus, our search sensitivity is limited by the sensitivity of the existing reference data.
In fact, J-GEM24b was selected as a candidate at the beginning, but it is most likely a faint, red source, which is not detected in the reference {\Ks}-band data.
The situation of NIR deep imaging will be improved by survey with Rubin/LSST (up to $Y$-band, \cite{Ivezi2019ApJ...873..111I}) and Euclid (up to 2 $\mu$m, \cite{euclid}) and future survey with Roman (up to 2.3 $\mu$m, although the sky coverage is at most about 2000 deg$^2$).
For real-time deep NIR search for EM counterpart, it is important to prepare these reference images in advance to readily use for the transient search.

Another challenge specific for the case of \gw\ is the fact that the localization area largely overlaps with the Galactic plane.
Due to this fact, as a galaxy-targeted survey, our transient search suffers from contamination of possible Galactic objects.
In fact, the exact nature of several candidates remains unclear.
To effectively identify the nature of these Galactic objects, it is useful to visit the same field multiple times within the same night (ideally in the same filter), which is also helpful to remove moving objects in the Solar system.

Note that the expected detection rate of such Galactic objects is not well quantified.
For example, as estimated in \mbox{sub-subsection~\ref{sec:galactic}}, the event rate of DN outburst is still highly uncertain \citep{rau07}. 
Up-coming time-domain survey with Rubin/LSST will provide a deeper catalog of Galactic transients.
Furthermore, the deep survey will also enable to measure the detection rate and their statistical properties (such as amplitude and duration) of Galactic transients, which can be used to estimate the chance coincidence probability of contaminating objects.
Based on this information, we can significantly improve the efficiency of identifying EM counterparts of GW sources.

\section{Conclusions}
\label{sec:conclusions}
In this paper, we present the results of NIR follow-up observations of the GW event \gw\ using the Subaru Telescope/MOIRCS. \gw\ was initially classified as a BH–NS merger (with $> 99\%$ probability), with a strong implication of possible EM emission.
We started follow-up observations approximately 7.8 hours after the event.
Over two nights, we observed $\numofmoircsobsgalaxies$ nearby galaxies in the NIR $Y$ and {\Ks} bands, with a typical $3 \sigma$ limiting magnitude of around 21.4 and 21.1 AB mag, respectively. 
The targeted galaxies are selected based on the GLADE catalog and the 3D localization map: our survey corresponds to a completeness of \completeness\%.
Although later GW analyses reclassified \gw\ as likely terrestrial noise, our observations serve as a valuable test case for future NIR follow-up observations.

As a result of transient search, we find five EM counterpart candidates. 
Although none of the five candidates can be conclusively classified using only their photometric variability and color information, their behaviors are clearly inconsistent with that of a kilonova such as AT2017gfo.
Of the detected candidates, four exhibit observational characteristics that are consistent with known classes of transients other than kilonovae, such as SNe or DNe. 
In contrast, properties of a candidate J-GEM24a are not consistent with any known transient.
Its rapid fading, by more than five magnitudes in the {\Ks} band within six days, suggests the possibility of a previously unrecognized red and rapidly fading transient. 
Future large-scale, systematic NIR surveys may uncover similar objects and provide further insight into their nature.

We also demonstrate how our NIR follow-up observations can constrain kilonova models under the assumption of a non-detection.
NIR kilonova searches are particularly effective for events that are difficult to detect in optical, such as those viewed from the equatorial direction.
Also, NIR emission tends to evolve slowly and remains detectable for more than ten days.
Based on the non-detections at 0.4 and 1.4 days after the merger, we constrain the parameter space of the ejecta mass ($> 0.01 M_{\odot}$) and opacity ($< 5 \ {\rm cm^2\ g^{-1}}$).
Furthermore, assuming additional follow-up observations a few days later, we show that even higher-opacity regions of the parameter space can be constrained.
These results confirm that deep NIR observations offer an effective way to detect or constrain red kilonovae even at 200 Mpc.

Our observational results provide several important insights and challenges for future GW counterpart searches by NIR facilities such as WINTER, PRIME, Euclid, and Roman.
We find that one of the main limitations to survey sensitivity is the lack of sufficiently deep pre-existing reference images or catalogs, especially in the {\Ks} band: the shallow depth of available data made candidate identification more difficult.
Another important challenge arises when the localization region overlaps with the Galactic plane.
In such cases, as seen for \gw\, contamination by Galactic transients such as DNe becomes a significant issue.
Multiple observations within a single night, enabling evaluation of variability, are effective for distinguishing such sources.
Furthermore, the event rates and statistical properties of Galactic transients remain uncertain, and future time-domain surveys by Rubin/LSST are expected to improve our understanding of these populations.
Accumulating such knowledge will be a key to enhance the success rate to identify EM counterparts to GW events.

\begin{ack}
We are grateful to Makoto Uemura for valuable comments that significantly improved the discussion section.

This research is based in part 
on data collected at the Subaru Telescope, which is operated by the National Astronomical Observatory of Japan. We are honored and grateful for the opportunity of observing the Universe from Maunakea, which has the cultural, historical, and natural significance in Hawaii.

This research is supported by the Optical and Infrared Synergetic Telescopes for Education and Research (OISTER) program funded by the MEXT of Japan.

The Pan-STARRS1 Surveys (PS1) and the PS1 public science archive have been made possible through contributions by the Institute for Astronomy, the University of Hawaii, the Pan-STARRS Project Office, the Max-Planck Society and its participating institutes, the Max Planck Institute for Astronomy, Heidelberg and the Max Planck Institute for Extraterrestrial Physics, Garching, The Johns Hopkins University, Durham University, the University of Edinburgh, the Queen's University Belfast, the Harvard-Smithsonian Center for Astrophysics, the Las Cumbres Observatory Global Telescope Network Incorporated, the National Central University of Taiwan, the Space Telescope Science Institute, the National Aeronautics and Space Administration under Grant No. NNX08AR22G issued through the Planetary Science Division of the NASA Science Mission Directorate, the National Science Foundation Grant No. AST–1238877, the University of Maryland, Eotvos Lorand University (ELTE), the Los Alamos National Laboratory, and the Gordon and Betty Moore Foundation.

Atlas Image obtained as part of the Two Micron All Sky Survey (2MASS), a joint project of the University of Massachusetts and the Infrared Processing and Analysis Center/California Institute of Technology, funded by the National Aeronautics and Space Administration and the National Science Foundation.

This research is partially based on observations collected at the European Organisation for Astronomical Research in the Southern Hemisphere under ESO programme 180.22JF.023 and the ESO public survey VVVX, Programme ID 198.B-2004.

The VISTA Hemisphere Survey data products served at Astro Data Lab are based on observations collected at the European Organisation for Astronomical Research in the Southern Hemisphere under ESO programme 179.A-2010, and/or data products created thereof.

The Dark Energy Camera Plane Survey (DECaPS; NOAO Proposal ID 2016A-0323 and 2016B-0279, PI: D. Finkbeiner) includes data obtained at the Blanco telescope, Cerro Tololo Inter-American Observatory, National Optical Astronomy Observatory (NOAO).

The NSF NOIRLab is operated by the Association of Universities for Research in Astronomy (AURA) under a cooperative agreement with the National Science Foundation. Database access and other data services are provided by the ASTRO Data Lab.

This project used data obtained from the Dark Energy Camera (DECam), which was constructed by the Dark Energy Survey (DES) collaboration. Funding for the DES Projects has been provided by the U.S. Department of Energy, the U.S. National Science Foundation, the Ministry of Science and Education of Spain, the Science and Technology Facilities Council of the United Kingdom, the Higher Education Funding Council for England, the National Center for Supercomputing Applications at the University of Illinois at Urbana--Champaign, the Kavli Institute of Cosmological Physics at the University of Chicago, the Center for Cosmology and Astro-Particle Physics at The Ohio State University, the Mitchell Institute for Fundamental Physics and Astronomy at Texas A\&M University, Financiadora de Estudos e Projetos, Funda\c{c}\~{a}o Carlos Chagas Filho de Amparo \`a Pesquisa do Estado do Rio de Janeiro, Conselho Nacional de Desenvolvimento Cient\'{\i}fico e Tecnol\'{o}gico and the Minist\'{e}rio da Ci\^encia, Tecnologia e Inova\c{c}\~{a}o, the Deutsche Forschungsgemeinschaft, and the Collaborating Institutions in the Dark Energy Survey.

The collaborating institutions are Argonne National Laboratory, the University of California at Santa Cruz, the University of Cambridge, Centro de Investigaciones Energ\'{e}ticas, Medioambientales y Tecnol\'{o}gicas--Madrid, the University of Chicago, University College London, the DES-Brazil Consortium, the University of Edinburgh, the Eidgen\"ossische Technische Hochschule (ETH) Z\"urich, Fermi National Accelerator Laboratory, the University of Illinois at Urbana-Champaign, the Institut de Ci\`{e}ncies de l'Espai (IEEC/CSIC), the Institut de F\'{\i}sica d'Altes Energies, Lawrence Berkeley National Laboratory, the Ludwig--Maximilians--Universit\"at M\"unchen and the associated Excellence Cluster Universe, the University of Michigan, the National Optical Astronomy Observatory, the University of Nottingham, The Ohio State University, the OzDES Membership Consortium, the University of Pennsylvania, the University of Portsmouth, SLAC National Accelerator Laboratory, Stanford University, the University of Sussex, and Texas A\&M University.

\end{ack}

\section*{Funding}
This work is partially supported by the Grant-in-Aid for Scientific research from JSPS (grant Nos. 21H04997, 23H00127, 23H04891, 23H04894, 23H05432, 24H00027, 24K07090), the JST FOREST Program (grant No. JPMJFR212Y).

\section*{Data availability} 
The data used in this research will be shared on reasonable request to the corresponding author.

\appendix 
\section{Detailed measurements of the candidates}
\label{sec:appendix_measurements}
Here we measure the shape and motion of each candidate transient to verify whether they are moving objects.

We first examine the shapes of the candidates as compared with those of surrounding stars.
The PSF size of the surrounding stars is determined by first constructing an average PSF from bright point sources and then fitting this average PSF with a two-dimensional Gaussian function to obtain the mean FWHMs along the X and Y directions.
The sizes of the EM counterpart candidates are also measured by fitting a two-dimensional Gaussian function.
The measured FWHMs are summarized in table~\ref{tab:comp_fwhm}.
The FWHM of J-GEM24f is slightly larger than that of the surrounding stars, but still consistent with the average PSF within 2$\sigma$.
The sizes of the other candidates/detections are fully consistent with the point sources.

\begin{figure}[htbp]
  \includegraphics[width=0.88\columnwidth]{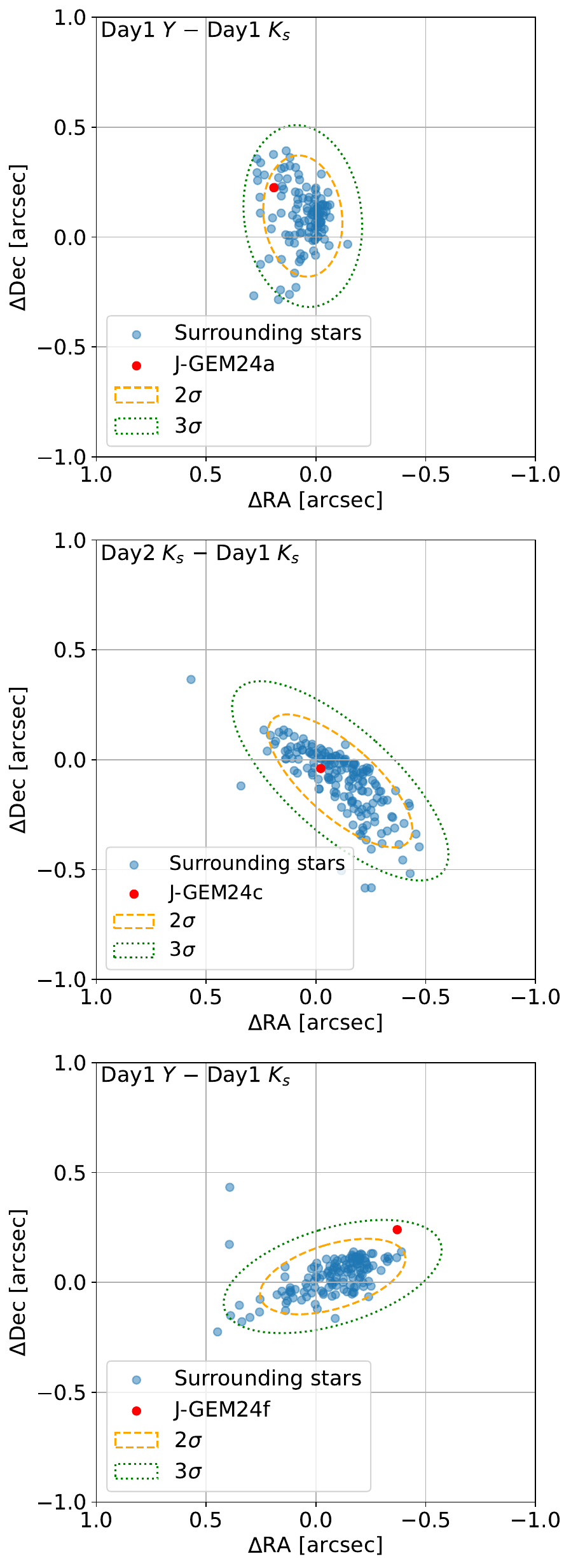}
  \caption{
    Positional offsets of J-GEM24a (top), J-GEM24c (middle), and J-GEM24f (bottom) between the Day~1 {\Ks}- and $Y$-band observations (for J-GEM24a and 24f) and Day~1 and Day 2 {\Ks} bands (for J-GEM24c).
    Blue points show surrounding stars, and the red point marks the candidate.
    Orange dashed and green dotted ellipses indicate the 2$\sigma$ and 3$\sigma$ confidence regions, respectively.
    {Alt text: Three-panel plot showing positional offsets of candidates relative to surrounding stars. Each panel shows one candidate point and surrounding star points. Confidence ellipses, derived from the distribution of surrounding stars, indicate 2-sigma and 3-sigma levels. Axes show shifts in right ascension and declination in arcseconds.}
    }
  \label{fig:astrometry}
\end{figure}

We then investigate the apparent motion of the candidates. 
This analysis focuses on J-GEM24a, 24c, and 24f, which are detected multiple times during the observation period.
For each candidate, positional offsets in right ascension (RA) and declination (Dec) are compared with the offset distribution of surrounding point sources located within 1.5 arcmin of the target position (the search radius is adjusted, when necessary, to include more than 100 stars in total).
For J-GEM24a and 24f, the offsets are measured as the positional differences between the Day~1 {\Ks}- and $Y$-band observations, whereas for J-GEM24c they are measured as the differences between the Day~1 and Day~2 {\Ks}-band observations.
Figure~\ref{fig:astrometry} presents the positional changes for 
J-GEM24a 24c, and 24f.
In all cases, the positional offsets of the candidates are consistent with those of the surrounding stars and lie within the 3$\sigma$ confidence region, indicating no significant motion during the observation period.

\begin{table}[t]
\tbl{FWHMs of EM counterpart candidates and surrounding stars.}{
\begin{tabular}{llrr}
\hline
candidate  & filter & FWHM$_{\rm cand}$ [\arcsec] &  FWHM$_{\rm av}$ [\arcsec]\\
\hline
J-GEM24a & $Y$   & $0.93 \pm 0.22$ & $1.02 \pm 0.01$\\
J-GEM24a & {\Ks}  & $1.16 \pm 0.61$ & $0.86 \pm 0.02$ \\
J-GEM24e & $Y$  & $0.98 \pm 0.20$ & $0.90 \pm 0.01$  \\
J-GEM24f & $Y$   & $1.55 \pm 0.36$  & $0.94 \pm 0.01$\\
J-GEM24f & {\Ks} & $1.73 \pm 0.52$  & $0.74 \pm 0.01$ \\
\hline
\end{tabular}
}\label{tab:comp_fwhm}
\begin{tabnote}
\hangindent6pt\noindent
All the measurements are performed for the data at Day~1.
\end{tabnote}
\end{table}
%

\section{List of observed galaxies}
\label{sec:appendix_gallist}
Table~\ref{tab:tab_gallist} provides a complete list of the GLADE galaxies observed with Subaru/MOIRCS during our follow-up observations.
The table includes the galaxy name, equatorial coordinates (J2000.0), luminosity distance $d_L$ (in Mpc), weighted probability $P^i$ (see \mbox{equation~\ref{eq:pi_prob}}), and the number of MOIRCS observations, in the following order: $Y$ band on Day~1, $K_s$ band on Day~1, $Y$ band on Day~2, and $K_s$ band on Day~2.

\setlength{\tabcolsep}{8pt}
\begin{longtable}{*{6}{l}}
\caption{
GLADE galaxies observed with MOIRCS.
}
\label{tab:tab_gallist} \\
\hline
      galaxy name & RA (J2000.0) [deg] & Dec (J2000.0) [deg] & $d_L$ [Mpc] & $P^i$ & $n_{\rm{obs}}$ \\
\hline
\endfirsthead
\hline
      galaxy name & RA (J2000.0) [deg] & Dec (J2000.0) [deg] & $d_L$ [Mpc] & $P^i$ & $n_{\rm{obs}}$ \\
\hline
\endhead
\hline
\endfoot
\hline
\endlastfoot
GL082154$-$300605 & 125.4736 & $-$30.1013 & 245 & 0.00029 & 0, 0, 1, 1 \\
GL082638$-$204958 & 126.6594 & $-$20.8328 & 214 & 0.00027 & 0, 0, 1, 1 \\
GL080327$-$260010 & 120.8631 & $-$26.0027 & 287 & 0.00003 & 0, 0, 1, 1 \\
GL074315$-$254550 & 115.8113 & $-$25.7639 & 126 & 0.00002 & 0, 0, 0, 1 \\
GL080313$-$241958 & 120.8045 & $-$24.3327 & 339 & 0.00000 & 0, 0, 1, 1 \\
GL075742$-$215313 & 119.4243 & $-$21.8870 & 216 & 0.00020 & 0, 0, 1, 1 \\
GL080605$-$222702 & 121.5212 & $-$22.4506 & 264 & 0.00009 & 0, 0, 0, 1 \\
GL075344$-$262712 & 118.4321 & $-$26.4532 & 164 & 0.00170 & 1, 2, 1, 1 \\
GL075620$-$312150 & 119.0820 & $-$31.3638 & 200 & 0.00100 & 1, 2, 0, 0 \\
GL075736$-$263533 & 119.4000 & $-$26.5924 & 211 & 0.00071 & 0, 0, 1, 1 \\
GL075745$-$312020 & 119.4365 & $-$31.3390 & 164 & 0.00080 & 0, 0, 1, 1 \\
GL075851$-$343144 & 119.7135 & $-$34.5289 & 184 & 0.00129 & 1, 2, 0, 0 \\
GL075900$-$241119 & 119.7467 & $-$24.1886 & 141 & 0.00075 & 0, 0, 1, 1 \\
GL075906$-$305442 & 119.7738 & $-$30.9117 & 170 & 0.00074 & 0, 0, 1, 1 \\
GL075925$-$273216 & 119.8544 & $-$27.5379 & 174 & 0.00120 & 1, 2, 0, 0 \\
GL075951$-$284307 & 119.9624 & $-$28.7187 & 142 & 0.00083 & 0, 0, 1, 1 \\
GL080004$-$274149 & 120.0171 & $-$27.6970 & 203 & 0.00079 & 0, 0, 1, 1 \\
GL080019$-$244809 & 120.0790 & $-$24.8024 & 199 & 0.00074 & 0, 0, 1, 1 \\
GL080037$-$262017 & 120.1550 & $-$26.3380 & 221 & 0.00077 & 0, 0, 1, 1 \\
GL080121$-$325937 & 120.3394 & $-$32.9937 & 180 & 0.00077 & 0, 0, 1, 1 \\
GL080138$-$294027 & 120.4096 & $-$29.6740 & 205 & 0.00085 & 0, 0, 1, 1 \\
GL080156$-$232917 & 120.4827 & $-$23.4882 & 220 & 0.00079 & 0, 0, 1, 1 \\
GL080201$-$305049 & 120.5031 & $-$30.8469 & 231 & 0.00153 & 1, 2, 0, 0 \\
GL080231$-$224547 & 120.6276 & $-$22.7630 & 162 & 0.00097 & 0, 0, 1, 1 \\
GL080236$-$340407 & 120.6480 & $-$34.0687 & 211 & 0.00100 & 1, 2, 0, 0 \\
GL080242$-$314115 & 120.6762 & $-$31.6875 & 190 & 0.00154 & 1, 2, 0, 0 \\
GL080305$-$224100 & 120.7727 & $-$22.6834 & 208 & 0.00084 & 0, 0, 1, 1 \\
GL080305$-$261201 & 120.7715 & $-$26.2002 & 177 & 0.00130 & 1, 2, 0, 0 \\
GL080332$-$223418 & 120.8822 & $-$22.5717 & 171 & 0.00078 & 0, 0, 1, 1 \\
GL080334$-$330027 & 120.8925 & $-$33.0074 & 222 & 0.00105 & 1, 2, 0, 0 \\
GL080339$-$330139 & 120.9107 & $-$33.0274 & 202 & 0.00117 & 1, 2, 0, 0 \\
GL080426$-$380023 & 121.1074 & $-$38.0063 & 160 & 0.00082 & 0, 0, 1, 1 \\
GL080436$-$250927 & 121.1483 & $-$25.1575 & 192 & 0.00102 & 1, 1, 0, 0 \\
GL080442$-$265813 & 121.1737 & $-$26.9703 & 158 & 0.00156 & 1, 2, 0, 0 \\
GL080444$-$313845 & 121.1848 & $-$31.6459 & 166 & 0.00081 & 0, 0, 1, 1 \\
GL080501$-$324313 & 121.2548 & $-$32.7203 & 148 & 0.00179 & 1, 2, 0, 0 \\
GL080513$-$281754 & 121.3057 & $-$28.2984 & 230 & 0.00075 & 0, 0, 1, 1 \\
GL080522$-$322313 & 121.3433 & $-$32.3869 & 215 & 0.00114 & 1, 2, 0, 0 \\
GL080528$-$315445 & 121.3678 & $-$31.9125 & 173 & 0.00104 & 1, 2, 0, 0 \\
GL080531$-$224000 & 121.3789 & $-$22.6668 & 202 & 0.00089 & 0, 0, 1, 1 \\
GL080542$-$315238 & 121.4270 & $-$31.8772 & 161 & 0.00148 & 1, 2, 0, 0 \\
GL080550$-$261139 & 121.4591 & $-$26.1942 & 181 & 0.00109 & 1, 2, 0, 0 \\
GL080551$-$255848 & 121.4613 & $-$25.9800 & 141 & 0.00211 & 1, 2, 0, 0 \\
GL080556$-$303009 & 121.4817 & $-$30.5026 & 236 & 0.00101 & 1, 2, 0, 0 \\
GL080604$-$222712 & 121.5179 & $-$22.4534 & 184 & 0.00078 & 0, 0, 1, 1 \\
GL080615$-$312154 & 121.5613 & $-$31.3651 & 173 & 0.00087 & 0, 0, 1, 1 \\
GL080620$-$312607 & 121.5841 & $-$31.4354 & 211 & 0.00194 & 1, 2, 0, 0 \\
GL080630$-$342242 & 121.6239 & $-$34.3783 & 191 & 0.00072 & 0, 0, 1, 1 \\
GL080633$-$300736 & 121.6358 & $-$30.1266 & 186 & 0.00145 & 1, 2, 0, 0 \\
GL080635$-$251122 & 121.6458 & $-$25.1895 & 218 & 0.00082 & 0, 0, 1, 1 \\
GL080640$-$305102 & 121.6656 & $-$30.8505 & 220 & 0.00070 & 0, 0, 1, 1 \\
GL080702$-$314302 & 121.7571 & $-$31.7173 & 221 & 0.00102 & 1, 2, 0, 0 \\
GL080708$-$290851 & 121.7829 & $-$29.1476 & 197 & 0.00138 & 1, 2, 0, 0 \\
GL080714$-$315729 & 121.8065 & $-$31.9582 & 142 & 0.00096 & 0, 0, 1, 1 \\
GL080720$-$244703 & 121.8345 & $-$24.7843 & 211 & 0.00078 & 0, 0, 1, 1 \\
GL080720$-$293536 & 121.8334 & $-$29.5935 & 148 & 0.00069 & 0, 0, 1, 1 \\
GL080724$-$264036 & 121.8482 & $-$26.6768 & 141 & 0.00103 & 1, 2, 0, 0 \\
GL080726$-$292734 & 121.8577 & $-$29.4596 & 188 & 0.00153 & 1, 2, 0, 0 \\
GL080747$-$275243 & 121.9464 & $-$27.8787 & 210 & 0.00102 & 1, 2, 0, 0 \\
GL080749$-$255717 & 121.9561 & $-$25.9548 & 158 & 0.00275 & 1, 1, 0, 0 \\
GL080750$-$290734 & 121.9573 & $-$29.1262 & 145 & 0.00161 & 1, 2, 0, 0 \\
GL080826$-$310301 & 122.1070 & $-$31.0502 & 191 & 0.00150 & 1, 2, 0, 0 \\
GL080847$-$305655 & 122.1957 & $-$30.9487 & 179 & 0.00074 & 0, 0, 1, 1 \\
GL080850$-$243120 & 122.2078 & $-$24.5223 & 237 & 0.00109 & 1, 1, 1, 1 \\
GL080855$-$300836 & 122.2275 & $-$30.1434 & 171 & 0.00114 & 1, 2, 0, 0 \\
GL080855$-$301552 & 122.2307 & $-$30.2646 & 217 & 0.00123 & 1, 2, 0, 0 \\
GL080903$-$242702 & 122.2606 & $-$24.4506 & 235 & 0.00127 & 1, 1, 0, 0 \\
GL080906$-$270633 & 122.2763 & $-$27.1092 & 174 & 0.00162 & 1, 2, 0, 0 \\
GL080912$-$243444 & 122.3015 & $-$24.5789 & 233 & 0.00084 & 0, 0, 1, 1 \\
GL080926$-$240558 & 122.3571 & $-$24.0995 & 153 & 0.00124 & 1, 1, 0, 0 \\
GL080938$-$303917 & 122.4092 & $-$30.6548 & 140 & 0.00084 & 0, 0, 1, 1 \\
GL081000$-$202700 & 122.4989 & $-$20.4499 & 207 & 0.00075 & 0, 0, 1, 1 \\
GL081024$-$302051 & 122.6005 & $-$30.3474 & 144 & 0.00092 & 0, 0, 1, 1 \\
GL081042$-$304908 & 122.6744 & $-$30.8188 & 137 & 0.00156 & 1, 2, 0, 0 \\
GL081051$-$330116 & 122.7131 & $-$33.0210 & 182 & 0.00203 & 1, 2, 0, 0 \\
GL081055$-$305129 & 122.7293 & $-$30.8580 & 181 & 0.00133 & 1, 2, 0, 0 \\
GL081106$-$234121 & 122.7751 & $-$23.6892 & 188 & 0.00119 & 1, 1, 0, 0 \\
GL081106$-$305231 & 122.7733 & $-$30.8752 & 168 & 0.00166 & 1, 2, 0, 0 \\
GL081116$-$305424 & 122.8183 & $-$30.9066 & 192 & 0.00207 & 1, 2, 0, 0 \\
GL081119$-$254058 & 122.8303 & $-$25.6829 & 168 & 0.00241 & 1, 1, 0, 0 \\
GL081123$-$321104 & 122.8469 & $-$32.1845 & 196 & 0.00102 & 1, 2, 0, 0 \\
GL081135$-$240855 & 122.8975 & $-$24.1487 & 252 & 0.00075 & 0, 0, 1, 1 \\
GL081135$-$314704 & 122.8966 & $-$31.7845 & 201 & 0.00072 & 0, 0, 1, 1 \\
GL081135$-$321943 & 122.8960 & $-$32.3286 & 131 & 0.00099 & 1, 2, 0, 0 \\
GL081137$-$290405 & 122.9058 & $-$29.0681 & 149 & 0.00089 & 0, 0, 1, 1 \\
GL081145$-$340542 & 122.9380 & $-$34.0950 & 159 & 0.00072 & 0, 0, 1, 1 \\
GL081155$-$235315 & 122.9782 & $-$23.8874 & 170 & 0.00099 & 1, 1, 0, 0 \\
GL081200$-$243339 & 122.9967 & $-$24.5607 & 248 & 0.00079 & 0, 0, 1, 1 \\
GL081205$-$312547 & 123.0193 & $-$31.4298 & 164 & 0.00096 & 0, 0, 1, 1 \\
GL081206$-$292357 & 123.0262 & $-$29.3991 & 208 & 0.00085 & 0, 0, 1, 1 \\
GL081214$-$290821 & 123.0584 & $-$29.1391 & 239 & 0.00071 & 0, 0, 1, 1 \\
GL081221$-$244709 & 123.0859 & $-$24.7858 & 184 & 0.00137 & 1, 1, 0, 0 \\
GL081225$-$271157 & 123.1044 & $-$27.1993 & 201 & 0.00092 & 0, 0, 1, 1 \\
GL081235$-$203457 & 123.1444 & $-$20.5825 & 153 & 0.00126 & 1, 1, 0, 0 \\
GL081239$-$315224 & 123.1611 & $-$31.8734 & 210 & 0.00116 & 1, 2, 1, 1 \\
GL081240$-$312406 & 123.1646 & $-$31.4017 & 172 & 0.00109 & 1, 2, 0, 0 \\
GL081248$-$313414 & 123.2007 & $-$31.5706 & 179 & 0.00102 & 1, 2, 0, 0 \\
GL081301$-$285304 & 123.2540 & $-$28.8843 & 228 & 0.00135 & 1, 2, 0, 0 \\
GL081311$-$310427 & 123.2964 & $-$31.0740 & 200 & 0.00079 & 0, 0, 1, 1 \\
GL081312$-$280235 & 123.2980 & $-$28.0429 & 199 & 0.00318 & 1, 2, 1, 1 \\
GL081317$-$262020 & 123.3211 & $-$26.3388 & 214 & 0.00086 & 0, 0, 1, 1 \\
GL081330$-$204609 & 123.3747 & $-$20.7692 & 180 & 0.00082 & 0, 0, 1, 1 \\
GL081338$-$205947 & 123.4073 & $-$20.9964 & 197 & 0.00121 & 1, 1, 0, 0 \\
GL081340$-$211703 & 123.4167 & $-$21.2842 & 166 & 0.00072 & 0, 0, 1, 1 \\
GL081351$-$323922 & 123.4643 & $-$32.6560 & 157 & 0.00080 & 0, 0, 1, 1 \\
GL081352$-$290731 & 123.4653 & $-$29.1253 & 198 & 0.00122 & 1, 2, 0, 0 \\
GL081401$-$243845 & 123.5041 & $-$24.6457 & 216 & 0.00082 & 0, 0, 1, 1 \\
GL081403$-$341622 & 123.5136 & $-$34.2727 & 202 & 0.00098 & 1, 2, 0, 0 \\
GL081406$-$262747 & 123.5235 & $-$26.4629 & 179 & 0.00090 & 0, 0, 1, 1 \\
GL081413$-$332820 & 123.5542 & $-$33.4722 & 173 & 0.00125 & 1, 2, 0, 0 \\
GL081417$-$311627 & 123.5729 & $-$31.2742 & 141 & 0.00084 & 0, 0, 1, 1 \\
GL081419$-$322317 & 123.5796 & $-$32.3880 & 206 & 0.00131 & 1, 2, 0, 0 \\
GL081425$-$322500 & 123.6033 & $-$32.4167 & 135 & 0.00123 & 1, 2, 0, 0 \\
GL081435$-$234301 & 123.6479 & $-$23.7169 & 173 & 0.00111 & 1, 1, 0, 0 \\
GL081435$-$331123 & 123.6467 & $-$33.1897 & 223 & 0.00071 & 0, 0, 1, 1 \\
GL081438$-$331902 & 123.6567 & $-$33.3172 & 191 & 0.00070 & 0, 0, 1, 1 \\
GL081450$-$224615 & 123.7091 & $-$22.7708 & 211 & 0.00114 & 1, 1, 0, 0 \\
GL081451$-$342315 & 123.7105 & $-$34.3874 & 235 & 0.00075 & 0, 0, 1, 1 \\
GL081452$-$292944 & 123.7146 & $-$29.4956 & 200 & 0.00236 & 1, 2, 0, 0 \\
GL081456$-$331857 & 123.7315 & $-$33.3158 & 136 & 0.00093 & 0, 0, 1, 1 \\
GL081502$-$225349 & 123.7584 & $-$22.8969 & 189 & 0.00168 & 1, 1, 0, 0 \\
GL081502$-$331829 & 123.7574 & $-$33.3080 & 196 & 0.00164 & 1, 2, 0, 0 \\
GL081503$-$331523 & 123.7624 & $-$33.2563 & 207 & 0.00084 & 0, 0, 1, 1 \\
GL081506$-$300753 & 123.7743 & $-$30.1313 & 226 & 0.00090 & 0, 0, 1, 1 \\
GL081514$-$235249 & 123.8079 & $-$23.8803 & 176 & 0.00146 & 1, 1, 0, 0 \\
GL081521$-$330708 & 123.8377 & $-$33.1189 & 199 & 0.00124 & 1, 2, 0, 0 \\
GL081529$-$304751 & 123.8719 & $-$30.7975 & 205 & 0.00115 & 1, 2, 0, 0 \\
GL081547$-$283019 & 123.9464 & $-$28.5053 & 180 & 0.00111 & 1, 1, 0, 0 \\
GL081549$-$233224 & 123.9522 & $-$23.5399 & 160 & 0.00132 & 1, 1, 0, 0 \\
GL081550$-$311104 & 123.9587 & $-$31.1845 & 243 & 0.00070 & 0, 0, 1, 1 \\
GL081551$-$224821 & 123.9605 & $-$22.8058 & 154 & 0.00177 & 1, 1, 0, 0 \\
GL081600$-$225336 & 124.0015 & $-$22.8933 & 140 & 0.00119 & 1, 1, 0, 0 \\
GL081608$-$315204 & 124.0342 & $-$31.8679 & 171 & 0.00198 & 1, 2, 0, 0 \\
GL081614$-$213024 & 124.0564 & $-$21.5066 & 213 & 0.00136 & 1, 1, 0, 0 \\
GL081623$-$233218 & 124.0977 & $-$23.5383 & 221 & 0.00070 & 0, 0, 1, 1 \\
GL081628$-$274822 & 124.1169 & $-$27.8062 & 181 & 0.00213 & 1, 1, 0, 0 \\
GL081634$-$203944 & 124.1401 & $-$20.6622 & 217 & 0.00151 & 1, 1, 0, 0 \\
GL081638$-$245039 & 124.1590 & $-$24.8443 & 124 & 0.00140 & 1, 1, 0, 0 \\
GL081645$-$203612 & 124.1895 & $-$20.6033 & 203 & 0.00115 & 1, 1, 0, 0 \\
GL081651$-$314535 & 124.2110 & $-$31.7598 & 211 & 0.00071 & 0, 0, 1, 1 \\
GL081658$-$255212 & 124.2429 & $-$25.8700 & 163 & 0.00136 & 1, 1, 0, 0 \\
GL081659$-$233123 & 124.2453 & $-$23.5230 & 207 & 0.00080 & 0, 0, 1, 1 \\
GL081703$-$234745 & 124.2611 & $-$23.7957 & 165 & 0.00120 & 1, 1, 0, 0 \\
GL081727$-$275926 & 124.3642 & $-$27.9905 & 116 & 0.00095 & 0, 0, 0, 1 \\
GL081728$-$315121 & 124.3685 & $-$31.8558 & 218 & 0.00079 & 0, 0, 1, 1 \\
GL081729$-$222931 & 124.3723 & $-$22.4919 & 215 & 0.00089 & 0, 0, 1, 1 \\
GL081737$-$311300 & 124.4038 & $-$31.2167 & 148 & 0.00186 & 1, 2, 0, 0 \\
GL081743$-$203614 & 124.4273 & $-$20.6039 & 164 & 0.00090 & 0, 0, 1, 1 \\
GL081751$-$260634 & 124.4633 & $-$26.1096 & 154 & 0.00118 & 1, 1, 0, 0 \\
GL081756$-$324248 & 124.4831 & $-$32.7134 & 155 & 0.00085 & 0, 0, 1, 1 \\
GL081801$-$161400 & 124.5025 & $-$16.2333 & 225 & 0.00090 & 0, 0, 1, 1 \\
GL081806$-$200632 & 124.5242 & $-$20.1089 & 190 & 0.00125 & 1, 1, 0, 0 \\
GL081827$-$260446 & 124.6126 & $-$26.0795 & 184 & 0.00074 & 0, 0, 1, 1 \\
GL081840$-$285532 & 124.6684 & $-$28.9256 & 188 & 0.00083 & 0, 0, 1, 1 \\
GL081842$-$263441 & 124.6764 & $-$26.5781 & 212 & 0.00081 & 0, 0, 1, 1 \\
GL081843$-$334427 & 124.6803 & $-$33.7409 & 147 & 0.00097 & 0, 0, 1, 1 \\
GL081854$-$324543 & 124.7258 & $-$32.7620 & 189 & 0.00120 & 1, 2, 2, 1 \\
GL081855$-$301427 & 124.7307 & $-$30.2409 & 164 & 0.00207 & 1, 2, 0, 0 \\
GL081859$-$231650 & 124.7445 & $-$23.2807 & 131 & 0.00149 & 1, 1, 0, 0 \\
GL081904$-$301803 & 124.7651 & $-$30.3009 & 134 & 0.00091 & 0, 0, 1, 1 \\
GL081904$-$370400 & 124.7650 & $-$37.0665 & 163 & 0.00134 & 1, 2, 0, 0 \\
GL081905$-$310416 & 124.7706 & $-$31.0712 & 179 & 0.00073 & 0, 0, 1, 1 \\
GL081907$-$255038 & 124.7804 & $-$25.8439 & 209 & 0.00089 & 0, 0, 1, 1 \\
GL081908$-$301502 & 124.7844 & $-$30.2506 & 211 & 0.00074 & 0, 0, 1, 1 \\
GL081914$-$284340 & 124.8093 & $-$28.7278 & 209 & 0.00098 & 1, 1, 0, 0 \\
GL081925$-$331610 & 124.8537 & $-$33.2694 & 209 & 0.00070 & 0, 0, 1, 1 \\
GL081943$-$322328 & 124.9292 & $-$32.3911 & 174 & 0.00087 & 0, 0, 1, 1 \\
GL081951$-$224338 & 124.9630 & $-$22.7273 & 127 & 0.00126 & 1, 1, 0, 0 \\
GL081952$-$211319 & 124.9658 & $-$21.2220 & 162 & 0.00091 & 0, 0, 1, 1 \\
GL082000$-$211900 & 124.9962 & $-$21.3167 & 158 & 0.00094 & 0, 0, 1, 1 \\
GL082008$-$195842 & 125.0323 & $-$19.9783 & 160 & 0.00074 & 0, 0, 1, 1 \\
GL082009$-$210955 & 125.0377 & $-$21.1652 & 127 & 0.00071 & 0, 0, 1, 1 \\
GL082029$-$183327 & 125.1212 & $-$18.5574 & 222 & 0.00073 & 0, 0, 1, 1 \\
GL082051$-$284958 & 125.2111 & $-$28.8328 & 177 & 0.00090 & 0, 0, 1, 1 \\
GL082052$-$281358 & 125.2148 & $-$28.2328 & 147 & 0.00134 & 1, 1, 0, 0 \\
GL082110$-$181309 & 125.2925 & $-$18.2191 & 222 & 0.00083 & 0, 0, 1, 1 \\
GL082115$-$281904 & 125.3126 & $-$28.3177 & 143 & 0.00094 & 0, 0, 0, 1 \\
GL082117$-$315346 & 125.3199 & $-$31.8961 & 151 & 0.00092 & 0, 0, 1, 1 \\
GL082119$-$292224 & 125.3290 & $-$29.3733 & 131 & 0.00138 & 1, 1, 0, 0 \\
GL082133$-$270940 & 125.3881 & $-$27.1611 & 163 & 0.00113 & 1, 1, 0, 0 \\
GL082144$-$221105 & 125.4313 & $-$22.1847 & 186 & 0.00109 & 1, 1, 0, 0 \\
GL082204$-$201701 & 125.5152 & $-$20.2836 & 174 & 0.00093 & 0, 0, 1, 1 \\
GL082208$-$254244 & 125.5316 & $-$25.7123 & 147 & 0.00073 & 0, 0, 1, 1 \\
GL082210$-$255316 & 125.5435 & $-$25.8878 & 194 & 0.00149 & 1, 1, 0, 0 \\
GL082214$-$204154 & 125.5598 & $-$20.6984 & 194 & 0.00093 & 0, 0, 1, 1 \\
GL082220$-$205627 & 125.5823 & $-$20.9409 & 177 & 0.00074 & 0, 0, 1, 1 \\
GL082220$-$285220 & 125.5832 & $-$28.8723 & 189 & 0.00087 & 0, 0, 1, 1 \\
GL082225$-$270438 & 125.6060 & $-$27.0773 & 184 & 0.00169 & 1, 1, 0, 0 \\
GL082242$-$324040 & 125.6761 & $-$32.6777 & 130 & 0.00080 & 0, 0, 1, 1 \\
GL082249$-$294256 & 125.7056 & $-$29.7155 & 206 & 0.00071 & 0, 0, 1, 1 \\
GL082314$-$253706 & 125.8095 & $-$25.6182 & 183 & 0.00090 & 0, 0, 1, 1 \\
GL082317$-$220507 & 125.8191 & $-$22.0852 & 155 & 0.00157 & 1, 1, 0, 0 \\
GL082342$-$291616 & 125.9261 & $-$29.2712 & 156 & 0.00179 & 1, 1, 0, 0 \\
GL082354$-$271106 & 125.9757 & $-$27.1851 & 164 & 0.00076 & 0, 0, 1, 1 \\
GL082416$-$315147 & 126.0683 & $-$31.8632 & 161 & 0.00097 & 0, 0, 1, 1 \\
GL082453$-$231312 & 126.2209 & $-$23.2201 & 213 & 0.00126 & 1, 1, 0, 0 \\
GL082551$-$261311 & 126.4612 & $-$26.2196 & 189 & 0.00094 & 0, 0, 1, 1 \\
GL082619$-$205052 & 126.5800 & $-$20.8478 & 185 & 0.00218 & 1, 1, 1, 1 \\
GL082626$-$272137 & 126.6090 & $-$27.3603 & 186 & 0.00175 & 1, 1, 0, 0 \\
GL082626$-$272138 & 126.6101 & $-$27.3604 & 190 & 0.00114 & 1, 1, 0, 0 \\
GL082658$-$155219 & 126.7434 & $-$15.8719 & 134 & 0.00090 & 0, 0, 1, 1 \\
GL082722$-$262347 & 126.8403 & $-$26.3965 & 186 & 0.00101 & 1, 1, 0, 0 \\
GL082731$-$175525 & 126.8788 & $-$17.9237 & 168 & 0.00095 & 0, 0, 1, 1 \\
GL082744$-$210647 & 126.9349 & $-$21.1131 & 207 & 0.00081 & 0, 0, 1, 1 \\
GL082847$-$274313 & 127.1972 & $-$27.7201 & 184 & 0.00088 & 0, 0, 1, 1 \\
GL083041$-$194242 & 127.6702 & $-$19.7118 & 155 & 0.00096 & 0, 0, 1, 1 \\

\end{longtable}

%

%

\end{document}